\documentclass[5p,authoryear]{elsarticle}

\usepackage[latin1]{inputenc}
\usepackage[T1]{fontenc}
\usepackage[english]{babel}
\usepackage{amsmath, amssymb}
\usepackage{hyperref}
\hypersetup{%
	final=true,%
	colorlinks=true%
}
\usepackage{hyphenat}
\usepackage[babel=true]{csquotes}
\usepackage[nohyperlinks,printonlyused]{acronym}
\usepackage{graphicx}
\usepackage{caption}
\usepackage{subcaption}
\usepackage{booktabs}
\usepackage{nicefrac}
\usepackage{tabularx}
\usepackage{ragged2e}
\usepackage{enumitem}

\usepackage{etoolbox}

\makeatletter

\patchcmd{\NAT@citex}
  {\@citea\NAT@hyper@{%
     \NAT@nmfmt{\NAT@nm}%
     \hyper@natlinkbreak{\NAT@aysep\NAT@spacechar}{\@citeb\@extra@b@citeb}%
     \NAT@date}}
  {\@citea\NAT@nmfmt{\NAT@nm}%
   \NAT@aysep\NAT@spacechar\NAT@hyper@{\NAT@date}}{}{}

\patchcmd{\NAT@citex}
  {\@citea\NAT@hyper@{%
     \NAT@nmfmt{\NAT@nm}%
     \hyper@natlinkbreak{\NAT@spacechar\NAT@@open\if*#1*\else#1\NAT@spacechar\fi}%
       {\@citeb\@extra@b@citeb}%
     \NAT@date}}
  {\@citea\NAT@nmfmt{\NAT@nm}%
   \NAT@spacechar\NAT@@open\if*#1*\else#1\NAT@spacechar\fi\NAT@hyper@{\NAT@date}}
  {}{}

\makeatother

\newcommand{\sdash}{\penalty10000---}
\newcommand{\edash}{\penalty10000---}

\newcommand{\eqnsref}[1]{Equations~\ref{#1}}

\newcommand{\tabsref}[1]{Tables~\ref{#1}}

\newcommand{\appref}[1]{\ref{#1}}

\newcommand{\RR}{\mathbb{R}}

\newcommand{\tn}[1]{\textnormal{#1}}

\DeclareMathOperator{\argmax}{arg\,max}

\newcommand*{\eg}{e.g.}

\newcommand*{\ie}{i.e.}
\newcommand*{\ibid}{ibid.}

\newcommand{\du}{d.u.}
\newcommand{\na}{n.a.}

\newcommand{\noparencite}[1]{\citeauthor{#1} \citeyear{#1}}

\newenvironment{paramTable}
	{\begin{table*} \centering \small}
	{\end{table*}}

\begin{document}
	
\acrodef{ACM}{Association for Computing Machinery}
\acrodef{ADR}{Advection\hyp{}Diffusion\hyp{}Reaction}
\acrodef{AMR}{Active Metabolic Rate}
\acrodef{ANOVA}{Analysis of Variance}
\acrodef{CG}{Conjugate Gradient}
\acrodef{CRS}{Compressed Row Storage}
\acrodef{DEVS}{Discrete Event System Specification}
\acrodef{DSL}{Domain\hyp{}Specific Language}
\acrodef{DoF}{Degree of Freedom}
\acrodefplural{DoF}[DoF]{Degrees of Freedom}
\acrodef{FCT}{Flux\hyp{}Corrected Transport}
\acrodef{FEM}{Finite Element Method}
\acrodef{FLOPS}{Floating Point Operations Per Second}
\acrodef{GPL}{General\hyp{}Purpose Language}
\acrodef{HPC}{High Performance Computing}
\acrodef{IBM}{Individual\hyp{}Based Model}
\acrodef{ICES}{International Council for the Exploration of the Sea}
\acrodef{IEEE}{Institute of Electrical and Electronics Engineers}
\acrodef{IPCC}{Intergovernmental Panel on Climate Change}
\acrodef{ISO}{International Organization for Standardization}
\acrodef{LOC}{Lines of Code}
\acrodef{LOWESS}{Locally Weighted Scatterplot Smoothing}
\acrodef{MDSE}{Model\hyp{}Driven Software Engineering}
\acrodef{MPI}{Message Passing Interface}
\acrodef{NAFO}{Northwest Atlantic Fisheries Organization}
\acrodef{NASA}{National Aeronautics and Space Administration}
\acrodef{NOAA}{National Oceanic and Atmospheric Administration}
\acrodef{NPZ}{Nutrient, Phytoplankton, Zooplankton}
\acrodef{ODE}{Ordinary Differential Equation}
\acrodef{PAR}{Photosynthetically Active Radiation}
\acrodef{PBE}{Population Balance Equation}
\acrodef{PDE}{Partial Differential Equation}
\acrodef{SI}{International System of Units}
\acrodef{SMR}{Standard Metabolic Rate}

\title{SPRAT: A Spatially\hyp{}Explicit Marine Ecosystem Model Based on\\Population Balance Equations}

\author[geomar,unikiel]{Arne N. Johanson\corref{cor1}}
\ead{arj@informatik.uni-kiel.de}
\author[geomar]{Andreas Oschlies}
\ead{aoschlies@geomar.de}
\author[unikiel]{Wilhelm Hasselbring}
\ead{wha@informatik.uni-kiel.de}
\author[dal]{Boris Worm}
\ead{bworm@dal.ca}
\cortext[cor1]{Corresponding author}
\address[geomar]{GEOMAR Helmholtz Centre for Ocean Research, D\"usternbrooker Weg 20, 24105 Kiel, Germany}
\address[unikiel]{Kiel University, Department of Computer Science, 24098 Kiel, Germany}
\address[dal]{Dalhousie University, Biology Department, Halifax, NS, Canada B3H 4R2}

\begin{abstract}
To successfully manage marine fisheries using an ecosystem\hyp{}based approach, long\hyp{}term predictions of fish stock development considering changing environmental conditions are necessary. 
Such predictions can be provided by end\hyp{}to\hyp{}end ecosystem models, which couple existing physical and biogeochemical ocean models with newly developed spatially\hyp{}explicit fish stock models. 
Typically, \acfp{IBM} and models based on \acf{ADR} equations are employed for the fish stock models.
In this paper, we present a novel fish stock model called SPRAT for end\hyp{}to\hyp{}end ecosystem modeling based on \acfp{PBE} that combines the advantages of \acp{IBM} and \ac{ADR} models while avoiding their main drawbacks. 
SPRAT accomplishes this by describing the modeled ecosystem processes from the perspective of individuals while still being based on partial differential equations.

We apply the SPRAT model to explore a well-documented regime shift observed on the eastern Scotian Shelf in the 1990s from a cod\hyp{}dominated to a herring\hyp{}dominated ecosystem. 
Model simulations are able to reconcile the observed multitrophic dynamics with documented changes in both fishing pressure and water temperature, followed by a predator\hyp{}prey reversal that may have impeded recovery of depleted cod stocks. 

We conclude that our model can be used to generate new hypotheses and test ideas about spatially interacting fish populations, and their joint responses to both environmental and fisheries forcing. 
\end{abstract}

\begin{keyword}
	end\hyp{}to\hyp{}end modeling \sep population balance equation \sep fish stock prediction \sep ecosystem\hyp{}based management
\end{keyword}

\maketitle

\renewcommand*{\sectionautorefname}{Section}
\renewcommand*{\subsectionautorefname}{Section}

\section{Introduction}

Living marine resources and their exploitation by fisheries play an important role in sustaining global nutrition but many of the world's fish stocks are in poor condition due to overharvesting \citep{worm2009,costello2016global}. 
This reduces the productivity of the stocks significantly and necessitates improved management in order to achieve a sustainable use of global fisheries resources. 

Fishing, however, is not the only impact on the condition and productivity of fish stocks but long- and short\hyp{}term variability of environmental parameters due to climate change or other sources of variability (such as the North Atlantic Oscillation (NAO)) imposes additional pressures \citep{brander2007}. 
The effects of changes in the environment on fish can be direct (\eg, by altering individual growth rates) or indirect (by affecting the net primary productivity and, thus, the carrying capacity of the ecosystem). 
Sometimes, these factors may interact with anthropogenic influences in complex ways. 
For example, the expansion of oxygen minimum zones in the tropical northeast Atlantic Ocean due to climate change compresses the suitable habitat of pelagic predator fish to a narrow surface layer and, thus, increases their vulnerability to surface fishing gear \citep{stramma2012}. 
The resulting high catch rates in such areas can lead to overly optimistic estimates of species abundance and, therefore, to exaggerated fishing quotas that put the affected stocks in danger of overexploitation. 

Another case illustrating the complexities of how fishing and climate can interact in driven rapid ecosystem change is the recent overfishing of Atlnatic cod (\textit{Gadus morhua}) stocks in the Gulf of Maine that occurred despite stringent management practises. 
Here, retrospective analysis showed that this change can in large part be attributed to rapid ocean warming that has led to an unrecognized effects on recruitment and mortality, and indirectly rendered fisheries exploitation rates unsustainable \citep{pershing2015slow}. 

From such examples, it becomes apparent that fisheries management must address the effects of fishing and climate variability and change in a joined framework, accounting for the effects of different sources of mortality, including changes in natural mortality, predation, and fishing \citep{rose2010}. 
Thus a more holistic ecosystem\hyp{}based approach has been called for, which may focus on marine ecosystems as a whole and takes into account the interdependence of their components \citep{cury2008}.

Ecological models that can supply this kind of information are sometimes called \emph{end\hyp{}to\hyp{}end models} because they incorporate all ecosystem components from the dynamics of the abiotic environment to primary producers to top predators \citep{travers2007}. 
In such models, the different elements of the ecosystem are linked together mainly through trophic interactions\sdash{}\ie, by feeding \citep{moloney2011}. 
Ideally, all these links between components are modeled bidirectionally (\eg, an increase in fish biomass due to feeding on zooplankton is also reflected in a decrease of zooplankton biomass). 
Such a two\hyp{}way coupling of model elements allows to explicitly resolve at the same time both bottom\hyp{}up and top\hyp{}down mechanisms of ecological control. 
It is the combination of modeling these bidirectional links in the trophic structure and considering the dynamics of the environment that enables end\hyp{}to\hyp{}end models to provide long\hyp{}term predictions on the development of fisheries ecosystems under environmental change. 
In the context of ecosystem\hyp{}based fisheries management, these predictive capabilities can be used to evaluate different management scenarios with regard to their long\hyp{}term effectiveness \citep{stock2011}. 

In practice, end\hyp{}to\hyp{}end models are typically constructed by using an existing physical and biogeochemical ocean model (for the abiotic environment as well as for nutrient and plankton dynamics) and creating a spatially\hyp{}explicit fish model that can be coupled with the ocean model \citep{shin2010}. 
In this context, fisheries are usually included in the model by assuming a mortality rate due to fishing (constant or changing with time), which applies homogeneously to the fish population beyond a certain lower size limit. 
Implementing a complete end\hyp{}to\hyp{}end model from scratch is discouraged by the amount of effort that is needed for developing sophisticated physical and biogeochemical models. 

The most widely used fish models for end\hyp{}to\hyp{}end modeling are either \acfp{IBM}, such as OSMOSE \citep{shin2001}, or models based on \acf{ADR} equations, such as SEA\-PODYM \citep{bertignac1998,lehodey2013}. 
\acp{IBM} offer the advantage that they are relatively easy to parametrize as their parameters are typically observable in individual fish. 
Additionally, these models can easily feature an emergent, dynamic food web structure. 
However, since\sdash{}at the ocean scale\edash{}it is not feasible to simulate all individual fish of the study region, so\hyp{}called \emph{super individual} approximations of \acp{IBM} are employed \citep{scheffer1995}. 
With this approach, individuals that share similar characteristics are replaced by a so\hyp{}called super individual\sdash{}\ie, an individual that has parameters similar to those of the individuals it represents plus an additional parameter that describes the number of individuals it stands for. 
This approximation technique is problematic because there is no mathematical framework for \acp{IBM} that would allow to formally study how many super\hyp{}individuals are necessary to simulate the inter\hyp{}individual interactions with sufficient accuracy. 

Since \ac{ADR} models are based on \acfp{PDE}, they integrate well with existing biogeochemical ocean models and feature a rigid mathematical framework with established approximation techniques for which formal error bounds can be described. 
Furthermore, \ac{ADR} equations are derived from the principle of mass conservation and are, thus, well\hyp{}suited for studying mass fluxes in marine ecosystems. 
However, \ac{ADR} models can be difficult to parametrize because most of their parameters are usually \emph{not} observable in individual fish and the full life cycle of the fish species is not directly represented in their main equation (only discrete age classes can be modeled). 

In order to combine the advantages of \acp{IBM} and \ac{ADR} models and to prevent their main drawbacks, we propose a fish model for end\hyp{}to\hyp{}end modeling that is based on so\hyp{}called \acfp{PBE} \citep{ramkrishna2000}. 
Our \ac{PBE} model\sdash{}which is called \emph{SPRAT}\edash{}represents fish as density distributions on a combined continuous space\hyp{}body size domain. 
Since \acp{PBE} are based on differential equations, they share the advantages of \ac{ADR} models with regard to the integration with existing biogeochemical ocean models and to the existence of established approximation techniques. 
At the same time, \ac{PBE} models share the distinct advantage of \acp{IBM} that most of their parameters can directly be observed in individual fish and that food web structure emerges dynamically from the model. 

Potential drawbacks introduced by our \ac{PBE}\hyp{}based mod\-el SPRAT in comparison to \acp{IBM} and \ac{ADR} models include:
\begin{enumerate}
	\item Since we represent fish as density distributions we cannot track fish and their interactions down to the level of single individuals (as it would be possible with an \ac{IBM} \emph{not} using the super individual approximation).
	\item In comparison to \ac{ADR} models, the SPRAT model is associated with increased computational costs because \ac{PBE} models represent the size of individuals as an additional dimension of the domain of a \ac{PDE}. 
\end{enumerate}
For a more detailed comparison of the \ac{PBE} approach with \acp{IBM} and \ac{ADR} models refer to \citet[Chap.~10]{johanson2016phd}. 

The \ac{PBE} approach to fish stock modeling is similar to so\hyp{}called size spectra models, which also describe fish via distribution functions on a continuous body size domain \citep[see, e.g.,][]{carozza2016,andersen2006,maury2013}. 
In the context of size spectra models, however, space is typically not resolved in the deduction of the models and is only introduced later on by assigning an instance of the respective model to each box or grid point of a discretized spatial grid (hence these models could be characterized as replicated one\hyp{}dimensional or univariate \ac{PBE} models). 
An exception to this is the APECOSM model by \citet{maury2010}, which is designed to study apex predators (namely tuna). 
APECOSM is a spatially\hyp{}continuous, mass\hyp{}balanced size spectrum model that, like SPRAT, offers a unified continuous description of fish in both space and body size via a single distribution function. 
Hence, SPRAT could also be described as a spatially\hyp{}continuous size spectrum model. 
Despite the strong similarities between these approaches, we prefer to call SPRAT a \ac{PBE} model to highlight that SPRAT is derived from a model type which is widely applied in engineering and has a large body of research associated with it \citep[especially regarding fast discretization techniques; see, e.g.,][]{le2016algorithms}.

In this paper, we apply SPRAT to simulate and mechanistically explore the complex interactions between the different components of the eastern Scotian Shelf ecosystem, specifically plankton and fish populatipons, fisheries and climate. 
The SPRAT model was implemented using a software engineering approach of the same name, which we presented earlier \citep{johanson2016sprat,johanson2014dada,johanson2014springsim}. 

\section{Material and Methods}

\subsection{Model Description}

Before describing the SPRAT model in detail, we first give a conceptual overview of the model, loosely following the ODD (Overview, Design concepts, Details) protocol by \citet{grimm2006}, which was designed as a standard protocol for describing \acp{IBM}. 

\begin{figure}
	\centering
	\includegraphics[width=\linewidth]{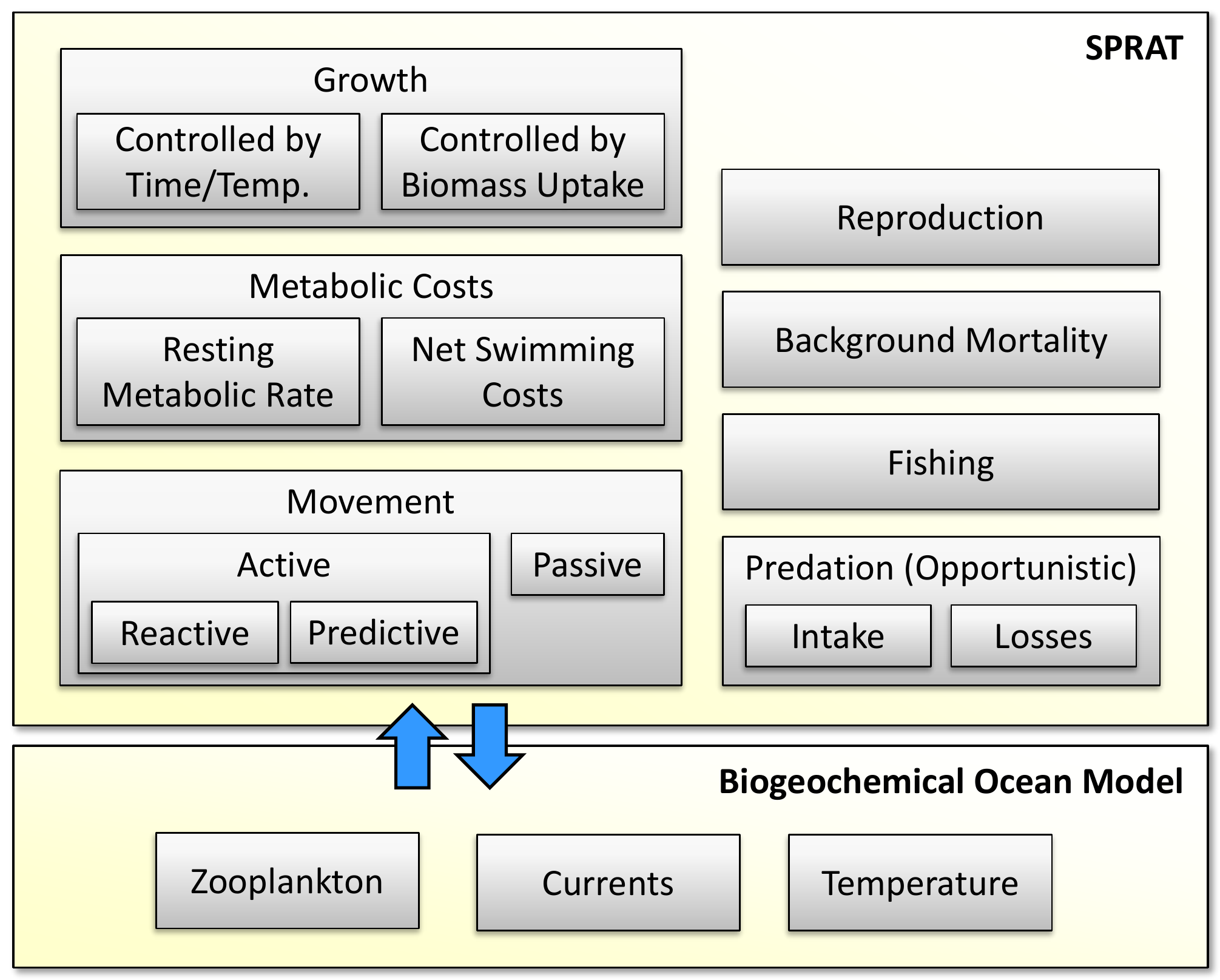}
	\caption{Conceptual diagram of the SPRAT model.}
	\label{fig:spratDiagram}
\end{figure}

\paragraph{Overview}
\begin{description}[leftmargin=0pt,font=\normalfont\itshape]
	\item[Purpose.] The main goal in the development of SPRAT is to construct a fish stock prediction model for end\hyp{}to\hyp{}end modeling which can represent fish in a completely continuous, rigorous mathematical framework and at the same time is still formulated from the perspective of the individual fish to allow for a dynamic food web structure. 
	The aim of this specific study is to employ SPRAT to simulate and explore a well-doumented regime shift on the eastern Scotian Shelf in the 1990s from a mainly cod\hyp{}dominated to a herring\hyp{}dominated ecosystem (\autoref{sec:scenario}). 
	\item[State variables and scales.] In our baseline simulation, we include five state variables: nutrient ($N$), phytoplankton ($P$), and zooplankton ($Z$) concentrations as well as mass distributions of two fish species complexes (predator and prey species representing mainly cod and herring). 
	All state variables vary in time and space (two spatial dimensions representing the vertically\hyp{}integrated ocean). 
	The two fish mass distributions furthermore vary in an additional body size dimension. 
	The $N$, $P$, and $Z$ variables are modeled discretely in space and continuously in time. 
	The fish mass distributions are modeled continuously in \emph{all} dimensions. 
	
	The vertically\hyp{}integrated ocean region of the eastern Scotian Shelf is assumed to be a homogeneous $450$ by $450$~km square. 
	For approximating the solution of our \ac{PDE}\hyp{}based model, this space is discretized into 48 by 48 equally\hyp{}sized rectangular cells. 
	The size dimension is divided into $32$ cells using logarithmically distributed division points. 
	The model is integrated in time from the year 1970 till 2010 with a time step of about 1 hour. 
	\item[Process overview.] An overview of the processes resolved by our model is given in \autoref{fig:spratDiagram}. 
	SPRAT is loosely based on dynamic energy budget models, which model the flow of energy or biomass through the ecosystem \citep[see, e.g.,][]{maury2013}. 
	The most important pathways in such models are predation, metabolic costs (resting metabolic rate and locomotion), structural growth, reproduction, and mortality (esp.\@ due to fishing). 
	Where applicable, we model these processes to be dependent on temperature because of the major significance of this variable for controlling metabolic rates \citep{clarke1999}. 
	In general, we include ocean currents into SPRAT but ignore them for the Scotian Shelf scenario because we make the simplifying assumption of space being homogeneous (see \autoref{sec:parametrization}). 
	
	Our fish stock model is coupled bi\hyp{}directionally with a simple, spatially\hyp{}explicit \acs{NPZ} model. 
	The \acs{NPZ} model resolves only one class of phytoplankton and zooplankton, each. 
	SPRAT and the \acs{NPZ} model are linked via the $Z$ state variable: when fish feed on zooplankton, the corresponding amount of biomass is transferred from the zooplankton state variable to the fish mass distribution. 
\end{description}

\paragraph{Design Concepts}
\begin{description}[leftmargin=0pt,font=\normalfont\itshape]
	\item[Mass conservation.] In the absence of sources and sinks, such as zooplankton grazing and fishing, the SPRAT model is mass\hyp{}conserving: the overall biomass in the system stays constant over time (the term \emph{\acl{PBE}} refers to the fact that the overall population mass may be distributed differently but, in total, stays in balance over time). 
	\item[Emergence.] The evolution of the fish density distributions is described from the perspective of the individual fish and no fixed food web structure is prescribed. 
	This implies that most of the parameters of SPRAT can be observed in individual fish and that the trophic structure of the ecosystem emerges dynamically from the model. 
	\item[Interaction.] In SPRAT, fish interact with each other mainly through feeding on each other. 
	In particular, the fish seek out locations, where the concentration of potential prey is high. 
	\item[Sensing.] We model fish to have perfect information about their environment within a certain radius. 
	In particular, we assume that they are able to assess, in which direction the maximum prey concentration lies within their radius of perfect information. 
	\item[Prediction.] If prey abundance levels fall below a certain level within the radius of perfect information, predatory fish will apply simple strategies (swim in a fixed direction) to find more suitable feeding grounds.
	\item[Stochasticity.] SPRAT is completely deterministic. 
\end{description}

\paragraph{Details}

In the SPRAT model, fish of species $\kappa = 1, \ldots, n$ are represented by the average carbon mass distribution 
\begin{equation} \label{eq:distribution}
m^{[\kappa]} \colon [0,t_{\max}] \times \Omega \to \RR , \, (t,x,y,r) \mapsto m^{[\kappa]}(t,x,y,r)
\end{equation}
with the combined space\hyp{}size domain 
\begin{equation}
\Omega = \Omega_S \times [ r_{\min},r_{\max} ] \text{.}
\end{equation}
Here, $t_{\max} \in \RR_{>0}$ describes an arbitrary time limit for the model, $\Omega_S$ is a two\hyp{}dimensional polygon domain representing the vertically integrated ocean, and $r_{\min}$ and $r_{\max}$ are the minimal and maximum carbon content masses of individual fish in the model, respectively. 
With carbon mass, we refer to the absolute mass of the carbon content of the dry mass of fish (both the mass distribution $m^{[\kappa]}$ and the body size dimension $r$ are carbon masses). 
By speaking of \emph{average} carbon mass, we mean that for any volume $V \subset \Omega$, the carbon mass of species $\kappa$ contained in that volume is given by $\int_V m^{[\kappa]}(t,x,y,r) \, d(x,y,r)$. 
The unit of $m^{[\kappa]}$ is kg C m$^{-2}$ (kg C)$^{-1}$ = m$^{-2}$. 
Note that the model uses a continuous size dimension instead of discrete size classes as often found in fish models used for end\hyp{}to\hyp{}end modeling purposes \citep[e.g.,][]{radtke2013,fulton2011,megrey2007}. 
\tabsref{tab:scotian_spratGlobalParameters} and \ref{tab:scotian_spratSpeciesParameters} in \appref{sec:parameterTables} provide an overview of all parameters of SPRAT. 

\begin{figure}
	\centering
	\includegraphics[width=0.65\linewidth]{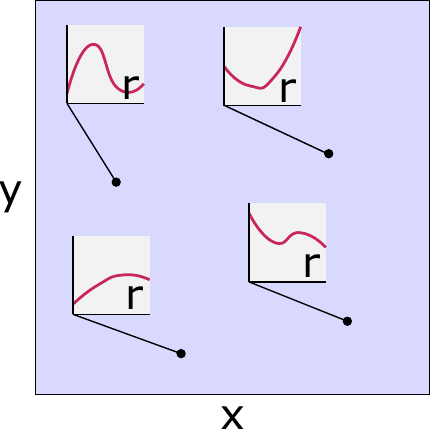}
	\caption{Conceptual visualization of the fish mass distributions in SPRAT. At each spatial point $(x,y)$, there is a one\hyp{}dimensional biomass distribution along the size dimension $r$.}
	\label{fig:distrVis}
\end{figure}

\autoref{fig:distrVis} provides a conceptual visualization of the distribution functions described by \autoref{eq:distribution}. 
For each fish species $\kappa$, at every point $(x,y)$ of the spatial domain (i.e., the vertically\hyp{}integrated ocean), there is a one\hyp{}dimensional distribution of fish biomass along the body size dimension. 
This one\hyp{}dimensional distribution describes how much mass $m^{[\kappa]} (t,x,y,r)$ of individuals of a certain mass $r$ there is in the system, at the current point in time $t$. 

Our model assumes constant ratios for carbon mass to dry mass ($C_{\tn{C/d}}^{[\kappa]}$) and dry mass to wet mass ($C_{\tn{d/w}}^{[\kappa]}$) for each species. 
For convenience, we define $M^{[\kappa]}$ to denote the wet mass distribution of species $\kappa$ and $u^{[\kappa]}$ to denote the average individual count density of the respective species. 

The evolution of the distributions $m^{[\kappa]}$ is governed by the following system of mass\hyp{}conserving population balance equations: 
\begin{equation} \label{eq:spratModel_popBal}
	\frac{\partial}{\partial t} m^{[\kappa]} + 
	\frac{\partial}{\partial x} q^{[\kappa]}_x m^{[\kappa]} + \frac{\partial}{\partial y} q^{[\kappa]}_y m^{[\kappa]} + \frac{\partial}{\partial r} g^{[\kappa]} m^{[\kappa]}
	= H^{[\kappa]}
\end{equation}
The vector $q^{[\kappa]} = ( q_x^{[\kappa]}, q_y^{[\kappa]} )$ is the spatial advection velocity of species $\kappa$ with unit m s$^{-1}$, $g^{[\kappa]}$ is a growth rate with unit kg C s$^{-1}$, and $H^{[\kappa]}$ is a source term with unit kg C m$^{-2}$ (kg C)$^{-1}$ s$^{-1}$ = m$^{-2}$ s$^{-1}$. 
The spatial derivatives ($\partial x$ and $\partial y$) describe the transport of fish in space with velocity $q^{[\kappa]}$ (due to both passive advection by currents and due to active locomotion). 
Correspondingly, the derivative in the size dimension ($\partial r$) describes growth with the dynamic growth rate $g^{[\kappa]}$. 
The source term $H^{[\kappa]}$ expresses how fish biomass is introduced to and removed from the system. 
In particular, the source term handles fishing as well as the mass re\hyp{}distribution occurring during feeding and reproduction. 

In the following sections, we specify the functional forms of the terms in \autoref{eq:spratModel_popBal} to include the concepts shown in the overview of the SPRAT model in \autoref{fig:spratDiagram}. 

\subsubsection{Length\hyp{}Weight Relationship}

In some contexts, we need the lengths of individuals instead of their weight. 
To convert wet mass $M$ (in kg) to length $l$ (in m), we use the power law 
\begin{equation} \label{eq:spratModel_weightLength}
M = a^{[\kappa]} l^{b^{[\kappa]}} \text{,}
\end{equation}
which implies that
\begin{equation} \label{eq:spratModel_lengthWeight}
l = \left( \frac{M}{a^{[\kappa]}} \right)^{\nicefrac{1}{b^{[\kappa]}}} \text{.}
\end{equation}

\subsubsection{Life Stages}

Although our model represents fish size using a continuous dimension, we need to differentiate between different life stages, each with its specific characteristics. 
For example, fish are only able to reproduce after they have reached maturity. 
Individuals transition from one life stage to the next once they grow past a certain size/mass. 

In the following, we list the masses that mark transitions relevant in our model: 
\begin{itemize}
	\item $w^{[\kappa]}_{\tn{egg}}$: dry weight of one egg from species $\kappa$
	\item $w^{[\kappa]}_{\tn{forage}}$: dry mass at which larvae of species $\kappa$ start to forage on their own (prior to that point, the larvae feed on their yolk sac)
	\item $M^{[\kappa]}_{\tn{plankton}}$: wet mass up to which individuals of species $\kappa$ consume zooplankton
	\item $M^{[\kappa]}_{\tn{mature}}$: wet mass at maturity of species $\kappa$
	\item $M^{[\kappa]}_{\tn{max}}$: maximum wet weight of an individual from species $\kappa$
\end{itemize}

\subsubsection{Spatial Velocity} \label{subsec:spratModel_movement}

The spatial transport velocity of fish from species $\kappa$ is given by: 
\begin{equation}
q^{[\kappa]} = \begin{cases}
V_{\tn{curr}} & \text{if } r < C_{\tn{C/d}}^{[\kappa]} w^{[\kappa]}_{\tn{forage}} \\
V_{\tn{curr}} + V_{\tn{react}}^{[\kappa]} + V_{\tn{pred}}^{[\kappa]} & \text{otherwise} 
\end{cases}
\end{equation}
For simplicity, we assume that fish which do not forage on their own have no active movement and are transported in space solely by currents ($V_{\tn{curr}}$; given by the ocean model SPRAT is coupled with). 
All other individuals are also affected by $V_{\tn{curr}}$ but exhibit two additional movement terms that together represent their active locomotion. 

In modeling the active movement of individuals, we follow the approach of \citet{ferno1998}, who describe fish swimming as being governed by two processes \citep[see also][]{neill1979}. 
The first process is \emph{reactive movement}, which means that the fish choose a swimming speed and direction based on their immediate surrounding. 
Thereby, they react to approaching predators, from which they flee, or to prey patches, by which they are attracted. 
The second process is \emph{predictive movement}, which is based not on stimuli from the current environment of the fish but rather on their experience and their instinct \citep[regarding learning in fish, see][]{brown2008}. 
Examples of this second type of movement are feeding and spawning migrations.

Reactive movement can be modeled based on a \emph{habitat index function}~$\mathfrak{H}^{[\kappa]}$, which describes how favorable a location is to an individual fish \citep[cf.\@][]{bertignac1998}. 
Using the index function, the fish can be modeled as trying to locally maximize $\mathfrak{H}^{[\kappa]}$ by swimming in a certain direction. 

In order to let the fish find the habitat that is locally optimal for them, it has been proposed \citep[\eg, by][]{bertignac1998} to set reactive movement velocity proportional to the (spatial) gradient of the habitat index.
We found that this optimization strategy is unsuitable for our model because it is too locally confined and produces large accumulations of fish in local maxima of $\mathfrak{H}^{[\kappa]}$ that are relatively inadequate compared to other close\hyp{}by locations. 
Therefore, we choose to assume that fish have perfect information about their surroundings within a certain radius $r_{\tn{view}}$ and swim to an absolute maximum of $\mathfrak{H}^{[\kappa]}$ within this area. 

The area of perfect information is given by 
\begin{equation} \label{eq:spratModel_informationSet}
	K(x,y) = \left\{ (v,w) \in \RR^2 : \lVert(v-x, w-y)\rVert \leq r_{\tn{view}} \right\} \cap \Omega_S \text{.}
\end{equation}
Within its bounds, we are looking for the spatial point with maximum habitat index and choose our swimming direction $\delta^{[\kappa]}_o (t,x,y,r)$ as 
\begin{equation} \label{eq:spratModel_swimDirection}
	\delta^{[\kappa]}_o = \left( \argmax_{(v,w) \in K(x,y)} \mathfrak{H}^{[\kappa]}(t,v,w,r) \right) - (x,y)
\end{equation}
with 
\begin{equation} \label{eq:spratModel_habitatIndex}
	\mathfrak{H}^{[\kappa]} = c^{[\kappa]}_{\tn{prey}} \text{.}
\end{equation}
The problem of finding an absolute maximum in $K(x,y)$ is computationally feasible since all densities are discretized for approximating the solution to \autoref{eq:spratModel_popBal} (hence, we need to check only a finite amount of points to find the desired maximum). 

Finally, we can define the reactive movement velocity as proportional to $\delta^{[\kappa]}_o$ as
\begin{equation}
V_{\tn{reactive}}^{[\kappa]} = \xi^{[\kappa]} \delta^{[\kappa]}_o
\end{equation}
with 
\begin{equation}
\xi^{[\kappa]} = \begin{cases}
\frac{\varsigma^{[\kappa]} l^{[\kappa]}(r)}{\lVert \delta^{[\kappa]}_o \rVert} & \text{if } \lVert \delta^{[\kappa]}_o \rVert > 0 \\
0 & \text{otherwise,}
\end{cases}
\end{equation}
where $\varsigma^{[\kappa]}$ is the cruise speed of fish from species $\kappa$ in body length per second.

The fish in our model employ predictive movement strategies in two cases: 
\begin{enumerate}
	\item If prey abundance is too low within the radius of perfect information $r_{\tn{view}}$ to sustain the fish, they initiate a feeding migration until they find more suitable habitat. 
	\item During the mating season, fish travel to spawning grounds. 
\end{enumerate}
Therefore, we define a migration velocity which is composed of feeding and spawning migration: 
\begin{equation}
V_{\tn{migr}}^{[\kappa]} = V_{\tn{feed}}^{[\kappa]} + V_{\tn{spawn}}^{[\kappa]}
\end{equation}

If there is no reactive movement (because no location within $r_{\tn{view}}$ has a higher habitat index than the current one) and there are more fish at the current location than can be sustained by the respective prey concentration, the fish have to migrate. 
This can be formalized as
\begin{equation} \label{eq:spratModel_Vfeeding}
	V_{\tn{feed}}^{[\kappa]} = \begin{cases}
		(1,0) & \text{if } \lVert V_{\tn{react}}^{[\kappa]} \rVert = 0 \land 
		m^{[\kappa]} > \zeta_0 c^{[\kappa]}_{\tn{prey}} \\
		0 & \text{otherwise,}
	\end{cases}
\end{equation}
where $\zeta_0$ is a linear scaling constant. 
The movement strategy we employ for feeding migrations is simple: 
swim in the direction of the positive $x$ axis (the swimming speed will be adjusted later on). 
A similarly simple strategy (\enquote{swim westwards}) has actually been observed in herring in the Norwegian Sea \citep{ferno1998}. 
Of course, more sophisticated strategies can be used, such as swimming to known feeding locations etc. 

For spawning migrations, we assume that there is a single spawning ground for each species given by its center $s^{[\kappa]} = (x^{[\kappa]}_s,y^{[\kappa]}_s)$. 
Furthermore, a fixed mating season is described by the subset $S^{[\kappa]} \subset [0,1]$, where the interval represents time of year (with $0$ being the first moment and $1$ the last moment of the year). 
We introduce the function $\theta(t) \in [0,1]$ to convert simulation time $t$ to time of year. 
With these prerequisites, the spawning migration velocity is given by 
\begin{equation}
	V_{\tn{spawn}}^{[\kappa]} = \begin{cases}
		(x^{[\kappa]}_s-x, y^{[\kappa]}_s-y) & \text{if } \theta(t) \in S^{[\kappa]} \\
		0 & \text{otherwise.}
	\end{cases}
\end{equation}
This approach can, of course, be extended to multiple spawning areas represented by regions rather than by points, if necessary. 

If there is a reason for the fish to migrate (\ie, $\lVert V_{\tn{migr}}^{[\kappa]} \rVert > 0$), we use the predictive movement velocity to combine migratory and reactive movement: 
\begin{equation}
	V_{\tn{pred}}^{[\kappa]} = \begin{cases}
		- V_{\tn{react}}^{[\kappa]} + & \\
		\; \varsigma^{[\kappa]} l^{[\kappa]}(r) 
		\frac{V_{\tn{react}}^{[\kappa]} + V_{\tn{migr}}^{[\kappa]}}{\lVert V_{\tn{react}}^{[\kappa]} + V_{\tn{migr}}^{[\kappa]} \rVert} & 
		\text{if } \lVert V_{\tn{migr}}^{[\kappa]} \rVert > 0 \, \land \\ & \hphantom{\text{if }} \lVert V_{\tn{react}}^{[\kappa]} + V_{\tn{migr}}^{[\kappa]} \rVert > 0 \\
		0 & \text{otherwise}
	\end{cases}
\end{equation}

\subsubsection{Sources and Sinks} \label{subsec:spratModel_forcings}

The source term of the model is given by 
\begin{equation}
	H^{[\kappa]} = \begin{cases}
		- m^{[\kappa]} \left( H_{\tn{fish}}^{[\kappa]} + H_{\tn{bg}}^{[\kappa]} \right) + H_{\tn{repr}}^{[\kappa]} & \\
		\quad \; + E^{[\kappa]} I^{[\kappa]} - L^{[\kappa]} - R^{[\kappa]} & \hspace*{-16pt}\text{if } r \geq C_{\tn{C/d}}^{[\kappa]} w^{[\kappa]}_{\tn{forage}} \\
		- m^{[\kappa]} H_{\tn{bg}}^{[\kappa]} + H_{\tn{repr}}^{[\kappa]} - L^{[\kappa]} & \hspace*{-16pt}\text{otherwise.}
	\end{cases}
\end{equation}
For fish which are able to actively forage, the source term consists of losses due to fishing ($H_{\tn{fish}}^{[\kappa]}$), background mortality ($H_{\tn{bg}}^{[\kappa]}$), predation ($L^{[\kappa]}$), and respiratory costs $R^{[\kappa]}$. 
The intake of food from predation is described by the term $E^{[\kappa]} I^{[\kappa]}$, where $I^{[\kappa]}$ is the gross carbon mass intake and $E^{[\kappa]}$ is the assimilation efficiency. 
The redistribution of mass during reproduction (from mature individuals to eggs) is covered by $H_{\tn{repr}}^{[\kappa]}$. 

For very early life stages that cannot forage on their own, only background mortality, reproduction, and losses from predation have to be considered. 

\paragraph{Fishing}
Extraction of individuals due to fishing is described by an instantaneous fishing mortality rate which can vary depending on spatial location and fish size. 
For example, for the Scotian Shelf scenario discussed in this paper, $H_{\tn{fish}}^{[\kappa]}$ is given by 
\begin{equation}
	H_{\tn{fish}}^{[\kappa]} (t,x,y,r) = \begin{cases}
		F^{[\kappa]}(t) & \text{if } M^{[\kappa]} \geq \frac{1}{2} M^{[\kappa]}_{\tn{mature}} \\
		0 & \text{otherwise,}
	\end{cases}
\end{equation}
where $F^{[\kappa]}(t)$ is the respective observed fishing mortality.

\paragraph{Background Mortality}
To account for losses of fish because of effects that are not explicitly considered in our model (such as predation by birds and marine mammals), we introduce the background mortality 
\begin{equation}
	H_{\tn{bg}}^{[\kappa]} = \begin{cases}
		\varepsilon^{[\kappa]}_{B} m^{[\kappa]} & \text{if } M^{[\kappa]} > \frac{1}{4} M^{[\kappa]}_{\tn{mature}} \\
		\varepsilon^{[\kappa]}_{B} m^{[\kappa]} - \zeta_B^{[\kappa]} \min (0, \Delta T(t)) & \text{otherwise.}
	\end{cases}
\end{equation}
For all individual sizes, we apply a quadratic (with respect to mass) instantaneous death term with mortality rate $\varepsilon^{[\kappa]}_{B}$. 
Since earlier life stages are especially vulnerable to fluctuations in temperature, we include an additional linear death term for smaller individuals that depends on the deviation $\Delta T(t)$ from the average habitat temperature \citep[see][]{houde2009}.

\paragraph{Metabolic Costs}
Respiratory costs are given by 
\begin{equation}
R^{[\kappa]} = \frac{m^{[\kappa]}}{r} \left( R_S^{[\kappa]} + R_A^{[\kappa]} \right) = u^{[\kappa]} \left( R_S^{[\kappa]} + R_A^{[\kappa]} \right) \text{,}
\end{equation}
where $R_S^{[\kappa]}$ is the \acf{SMR} of an individual fish and $R_A^{[\kappa]}$ represents the additional respiratory costs due to swimming (net swimming costs). 
$R_A^{[\kappa]}$ corresponds to the \acf{AMR} minus the \ac{SMR}. 

According to \citet{clarke1999}, \ac{SMR} is fitted well for many teleostei by 
\begin{equation}
\tn{SMR} = \frac{1}{5.43} \, M^{0.8}
\end{equation}
with $M$ being the wet mass of an individual in g and $\tn{SMR}$ in mmol O$_2$ h$^{-1}$. 

In order to convert oxygen consumption to carbon losses, we make use of the so\hyp{}called respiratory quotient $RQ$, which describes the ratio of exhaled CO$_2$ volume to inhaled O$_2$ volume: 
\begin{equation}
RQ = \frac{1 \tn{ mol CO}_2}{1 \tn{ mol O}_2}
\end{equation}
While the value of $RQ$ depends on the diet, \citet{videler1993} determines 
\begin{equation}
RQ = 0.96
\end{equation}
to be a good average value for fish. 
Therefore, it holds that 
\begin{equation} \label{eq:spratModel_RQ}
	1 \tn{ mmol O}_2\tn{ h}^{-1} = \frac{12}{0.96 \cdot 10^6 \cdot 60^2} \tn{ kg C s}^{-1} \text{.}
\end{equation}

By converting to appropriate units and compensating for temperature changes using a $Q_{10}$ temperature coefficient, we define the \ac{SMR} of an individual fish in our model as 
\begin{equation}
R_S^{[\kappa]} = \left( Q_{10}^{\tn{SMR}} \right)^{\frac{\Delta T(t)}{10}} 
\frac{12}{5.43 \cdot 0.96 \cdot 10^6 \cdot 60^2} \left( 10^3 M^{[\kappa]} \right)^{0.8} \text{.}
\end{equation}
\citet{videler1993} deems $Q_{10}^{\tn{SMR}} = 2$ to be appropriate in the context of resting metabolic rates of fish.

Regarding the respiratory net costs of swimming, \citet{boisclair1993} as well as \citet{ohlberger2005} propose a model of the form 
\begin{equation} \label{eq:spratModel_AMRModel}
	\tn{AMR} = \tn{SMR} + a M^b v^c \text{,}
\end{equation}
where $M$ is wet mass in g, $v$ is swimming speed in cm s$^{-1}$, and $\tn{AMR}$ is given in mg O$_2$ h$^{-1}$. 
A parametrization of \autoref{eq:spratModel_AMRModel} for steady swimming in many fish is $a = 10^{-2.43}$, $b = 0.8$, and $c = 1.21$ \citep{boisclair1993}. 
For determining $v$, we only have to consider the velocity due to active movement 
\begin{equation}
	V_{\tn{active}}^{[\kappa]} = V_{\tn{react}}^{[\kappa]} + V_{\tn{pred}}^{[\kappa]} \text{.}
\end{equation}
Applying appropriate unit conversions, we can define the net swimming costs per individual as: 
\begin{equation}
	R_A^{[\kappa]} = \frac{12 \cdot 10^{-2.43}}{0.96 \cdot 32 \cdot 10^6 \cdot 60^2} \left(10^3 M^{[\kappa]}\right)^{0.8} \left(10^2 \lVert V_{\tn{active}}^{[\kappa]} \rVert \right)^{1.21}
\end{equation}

\paragraph{Reproduction}
During the spawning season $S^{[\kappa]} \subset [0,1]$, mass is transferred from mature female fish to (fertilized) eggs. 
We assume a constant $50$~\%~:~$50$~\% sex ratio of females to males in the population. 
The net wet mass fecundity $\Phi^{[\kappa]}$ describes how many fertilized eggs per kg wet mass a mature female produces during the spawning season. 

In order to describe the mass transfer process related to reproduction, we define the carbon mass at maturity 
\begin{equation}
	r^{[\kappa]}_{\tn{mature}} = c_{\tn{C/d}}^{[\kappa]} c_{\tn{d/w}}^{[\kappa]} M^{[\kappa]}_{\tn{mature}} \text{,}
\end{equation}
the carbon mass of an egg 
\begin{equation}
r^{[\kappa]}_{\tn{egg}} = c_{\tn{C/d}}^{[\kappa]} w^{[\kappa]}_{\tn{egg}} \text{,}
\end{equation}
and the duration of the spawning season of species $\kappa$ in seconds 
\begin{equation}
	\theta^{[\kappa]}_{s} = 365 \cdot 24 \cdot 60^2 \cdot \int_{0}^{1} \mathfrak{1}_{S^{[\kappa]}} (\phi) \; d\phi \text{,}
\end{equation}
where $\mathfrak{1}_{S^{[\kappa]}}$ is the characteristic function on $S^{[\kappa]}$. 
This allows us to define the biomass spawning rate
\begin{equation}
	B^{[\kappa]}_{s} = \begin{cases}
		\frac{1}{2}
		\frac{\Phi^{[\kappa]} r^{[\kappa]}_{\tn{egg}}}{c_{\tn{C/d}}^{[\kappa]} c_{\tn{d/w}}^{[\kappa]} \theta^{[\kappa]}_{s}}
		m^{[\kappa]} & \text{if } \theta(t) \in S^{[\kappa]} \land r \geq r^{[\kappa]}_{\tn{mature}} \\
		0 & \text{otherwise,}
	\end{cases}
\end{equation}
which describes the amount of egg biomass that is spawned at each moment of the spawning season (spawning is assumed to happen at a constant rate during the whole season). 

In order to reintroduce the egg biomass into the distribution $m$ at an appropriate \enquote{place} in the size dimension, we define an insertion distribution $\psi^{[\kappa]}(r)$ that only has a small support which is centered around $r^{[\kappa]}_{\tn{egg}}$. 
We require $\psi^{[\kappa]} \colon [r_{\min},r_{\max}] \to \RR_{\geq 0}$ to be continuous and to fulfill 
\begin{align}
	&\argmax_{\phi \in [r_{\min},r_{\max}]} \left( \psi^{[\kappa]}(\phi) \right) = r^{[\kappa]}_{\tn{egg}} \\
	\land \; 
	&\int_{r_{\min}}^{r_{\max}} \psi^{[\kappa]}(\phi) \; d\phi = 1 \\
	\land \; 
	& \psi^{[\kappa]}(r) = 0 \text{ for all } r \geq r^{[\kappa]}_{\tn{mature}} \text{.}
\end{align}

The insertion distribution $\psi^{[\kappa]}(r)$ allows us to describe the redistribution of egg mass $H_{\tn{repr}}^{[\kappa]}(t,x,y,r)$ as 
\begin{equation} \label{eq:spratModel_reproduction}
	H_{\tn{repr}}^{[\kappa]} = \psi^{[\kappa]}(r) \int_{r^{[\kappa]}_{\tn{mature}}}^{r_{\max}} B^{[\kappa]}_{s} (t,x,y,\phi) \; d\phi
	- B^{[\kappa]}_{s} (t,x,y,r) \text{.}
\end{equation}

\subsubsection{Predation}

In this section, we describe the processes related to predation that determine the intake of biomass ($I^{[\kappa]}$), biomass losses due to predation ($L^{[\kappa]}$), and the grazing of zooplankton by fish ($G$). 
A key feature of marine food webs is opportunistic predation based on body size: 
fish tend to feed on organisms of all taxa, provided that the predator fish are physically able to ingest these organisms \citep{shin2010}. 
Therefore, in ecological models of fish, it is typically assumed that a predator can prey on other fish as soon as the ratio of predator to prey body length is larger than the \emph{minimal predator\hyp{}prey mass ratio} $\tau$ \citep{shin2001}. 

The amount of prey of species $\kappa$ that is available to a predator is given by 
\begin{equation}
S^{[\kappa]}_{p}(t,x,y,r) = \int_{r_{\min}}^{\nicefrac{r}{\tau}} m^{[\kappa]}(t,x,y,\phi) \; d\phi \text{.}
\end{equation}
Based on this, we define the auxiliary term 
\begin{align}
	F_m^{[\kappa]}(t,x,y,r) &= \eta^{[\kappa]} 
	\left( Q_{10}^{P} \right)^{\frac{\Delta T(t)}{10}}
	c_{\tn{C/d}}^{[\kappa]} c_{\tn{d/w}}^{[\kappa]} \\
	&\quad \cdot \left( M^{[\kappa]} \right)^{\alpha} 
	u^{[\kappa]} \label{eq:spratModel_predTemp} \frac{1}{F_0^{[\kappa]} + \sum_{i\neq \kappa} S_p^{[i]}}
\end{align}
to describe biomass intake as 
\begin{equation}
I^{[\kappa]} = G^{[\kappa]} + \begin{cases}
F_m^{[\kappa]} 
\sum_{i \neq \kappa} S_p^{[i]} & \text{if }  r \geq C_{\tn{C/d}}^{[\kappa]} w^{[\kappa]}_{\tn{forage}} \\
0 & \text{otherwise. }
\end{cases}
\end{equation}
Biomass intake of fish that forage on their own consists of zooplankton grazing $G^{[\kappa]}$ (see below) and predation on fish from other species (in its current form, the model does not take into account cannibalism although it could be included). 
Predation on fish is modeled via a saturation response with the half saturation constant $F_0^{[\kappa]}$. 
$\eta^{[\kappa]}$ is the constant predation rate of species $\kappa$ that is adjusted for temperature fluctuations by the temperature coefficient $Q_{10}^{P} = 2$. 
The allometric exponent $\alpha$ scales the influence of body mass on the predation rate and can be estimated to be about $0.74$ for pelagic predator fish \citep{ware1978}. 

The biomass losses due to predation are given by 
\begin{equation}
	L^{[\kappa]} = m^{[\kappa]} 
	\sum_{i\neq \kappa} \int_{\min(\max(\tau r, \, r^{[i]}_{\tn{forage}}), \, r_{\max})}^{r_{\max}} F_m^{[i]} (t,x,y,\phi) \; d\phi
\end{equation}
with $r^{[i]}_{\tn{forage}} = C_{\tn{C/d}}^{[i]} w^{[i]}_{\tn{forage}}$.
The complex expression for the lower integral boundary ensures that only predation by fish that are large enough ($\tau r$) and can forage on their own ($r^{[i]}_{\tn{forage}}$) is considered. 
Additionally, the lower integral boundary must not exceed the overall maximum fish size $r_{\max}$.

Regarding zooplankton grazing, we define an auxiliary term analogous to $F_m^{[\kappa]}$: 
\begin{equation} \label{eq:spratModel_grazingTemp}
	Z_m^{[\kappa]} = \mu^{[\kappa]} 
	\left( Q_{10}^{P} \right)^{\frac{\Delta T(t)}{10}}
	c_{\tn{C/d}}^{[\kappa]} c_{\tn{d/w}}^{[\kappa]} \left( M^{[\kappa]} \right)^{\alpha} 
	u^{[\kappa]} \text{,}
\end{equation}
where $\mu^{[\kappa]}$ is the zooplankton grazing rate of species $\kappa$. 
Using it, we characterize zooplankton grazing by a saturating response 
\begin{equation}
G^{[\kappa]} = \begin{cases}
Z_m^{[\kappa]} \cdot \frac{K_Z(t,x,y) \cdot c_Z(t,x,y)}{Z_0^{[\kappa]} + c_Z(t,x,y)} & \text{if } r \geq C_{\tn{C/d}}^{[\kappa]} w^{[\kappa]}_{\tn{forage}} \, \land \\
& \hphantom{\text{if }} M^{[\kappa]} \leq M^{[\kappa]}_{\tn{plankton}} \\
0 & \text{otherwise}
\end{cases}
\end{equation}
with $K_Z$ being a limiting factor, $Z_0^{[\kappa]}$ the half saturation constant, and $c_Z$ the local carbon mass concentration of zooplankton supplied by the biogeochemical model that SPRAT is coupled with.
The limiting factor $K_Z$ ensures that zooplankton cannot be depleted at arbitrarily high rates by depending on a linear zooplankton mortality rate ($j_f(Z)$) given by our \acf{NPZ} model, which is described in \appref{sec:appendixNPZ}. 
$K_Z$ is defined as 
\begin{equation} \label{eq:spratModel_ZLimiter}
K_Z(t,x,y) = \min \left( 1, \frac{\rho \cdot j_f(Z(t,x,y)) \cdot c_Z(t,x,y)}{k (t,x,y) + \varepsilon_Z} \right) \text{,}
\end{equation}
where $\rho$, $j_f$, and $Z$ are from our \ac{NPZ} model (see \appref{sec:appendixNPZ}). $\varepsilon_Z$ can be set to prevent division by zero, and $k$ is the (only theoretical) total unlimited zooplankton consumption by all species (compare the definitions of $G_{UL}^{[\kappa]}$ and $G^{[\kappa]}$): 
\begin{align}
k (t,x,y) &= \sum_{\kappa=1}^{n} \int_{r_{\min}}^{r_{\max}} G_{UL}^{[\kappa]} (t,x,y,\phi) \; d\phi \label{eq:spratModel_k} \\
G_{UL}^{[\kappa]} &= \begin{cases}
Z_m^{[\kappa]} \cdot \frac{c_Z(t,x,y)}{Z_0^{[\kappa]} + c_Z(t,x,y)} & \text{if } r \geq C_{\tn{C/d}}^{[\kappa]} w^{[\kappa]}_{\tn{forage}} \, \land \\
& \hphantom{\text{if }} M^{[\kappa]} \leq M^{[\kappa]}_{\tn{plankton}} \\
0 & \text{otherwise}
\end{cases} \label{eq:spratModel_GUL}
\end{align} 

In order to feed back the zooplankton consumption to the biogeochemical model, we define the total zooplankton consumption $G$: 
\begin{equation}
G(t,x,y) = \sum_{\kappa = 1}^{n} \int_{r_{\min}}^{r_{\max}} G^{[\kappa]} (t,x,y,\phi) \; d\phi
\end{equation}

\subsubsection{Growth} \label{subsec:spratModel_growth}

Somatic growth rate in kg s$^{-1}$ is given by 
\begin{equation}
	g^{[\kappa]} = \begin{cases}
		\left( Q_{10}^{G} \right)^{\frac{\Delta T(t)}{10}} \frac{\max\left(0, E^{[\kappa]} I^{[\kappa]}-R^{[\kappa]}\right)}{u^{[\kappa]}} & \text{if } \frac{w^{[\kappa]}_{\tn{forage}}}{c_{\tn{d/w}}^{[\kappa]}} \leq M^{[\kappa]}\\
		& \hphantom{if } < M^{[\kappa]}_{\tn{max}} \\
		\left( Q_{10}^{L} \right)^{\frac{\Delta T(t)}{10}} \frac{c_{\tn{C/d}}^{[\kappa]} \left( w_{\tn{forage}}^{[\kappa]} - w_{\tn{egg}}^{[\kappa]} \right)}{\theta_{\tn{yolk}}^{[\kappa]}} & \text{if } M^{[\kappa]} < \frac{w^{[\kappa]}_{\tn{forage}}}{c_{\tn{d/w}}^{[\kappa]}} \\
		0 & \text{otherwise.}
	\end{cases}
\end{equation}
We distinguish three cases: 
\begin{enumerate}
	\item Fish that can forage on their own but have not yet reached their maximum weight. 
	\item Fish that cannot yet forage on their own.
	\item Fish that have reached the maximum weight of their species.
\end{enumerate}
In the first two cases, we compensate for temperature fluctuations by using $Q_{10}$ temperature coefficients. 
For larvae/eggs, we employ $Q_{10}^{L} = 3$ and for further\hyp{}developed individuals, $Q_{10}^{G} = 2$ \citep[see][]{houde2009}. 

In the first case of individuals that forage on their own, $g^{[\kappa]}$ describes the positive net biomass assimilation rate of all individuals at the corresponding coordinates divided by the number of individuals. 
Thereby, it expresses the biomass accumulation rate of a single individual. 
Since mass is the only attribute to track the developmental stage of individuals in our model, we can only allow for positive growth rates because, otherwise, adult fish could, for example, become larvae again (\ie, develop backwards in time). 

In the second case of eggs/larvae that cannot (yet) forage on their own, we apply a constant growth rate (aside from the temperature dependency) to let them grow from $w_{\tn{egg}}^{[\kappa]}$ to $w_{\tn{forage}}^{[\kappa]}$ in time $\theta_{\tn{yolk}}^{[\kappa]}$. 
A constant growth rate is assumed to be a reasonably good approximation of the exponential growth that fish experience during their very early life stages, in which they (exclusively) feed on their yolk sac. 

One could argue, however, that eggs and larvae cannot accumulate mass as long as they do not take it up from their environment but from their yolk sac (the weight of which is already included in $w_{\tn{egg}}^{[\kappa]}$). 
Hence, it should hold that $w_{\tn{egg}}^{[\kappa]} = w_{\tn{forage}}^{[\kappa]}$. 
Since in our model we have to represent the aging process of eggs/larvae via changes in their mass (the model does not have an age dimension), we solve this problem pragmatically by letting $w_{\tn{egg}}^{[\kappa]}$ represent the lower end of the egg weight spectrum and $w_{\tn{forage}}^{[\kappa]}$ the upper end. 

By transporting a constant amount of mass from $w_{\tn{egg}}^{[\kappa]}$ to $w_{\tn{forage}}^{[\kappa]}$, we also introduce a hidden death term since the same mass at $w_{\tn{egg}}^{[\kappa]}$ represents more individuals than at $w_{\tn{forage}}^{[\kappa]} > w_{\tn{egg}}^{[\kappa]}$. 
This, however, does not have an effect on the mass conservation properties of the model. 

Once the fish have reached their maximum mass in the third case of the definition of $g^{[\kappa]}$, growth ceases. 

\subsection{Evaluation Scenario: Eastern Scotian Shelf} \label{sec:scenario}

\begin{figure}
	\centering
	\includegraphics[width=\linewidth]{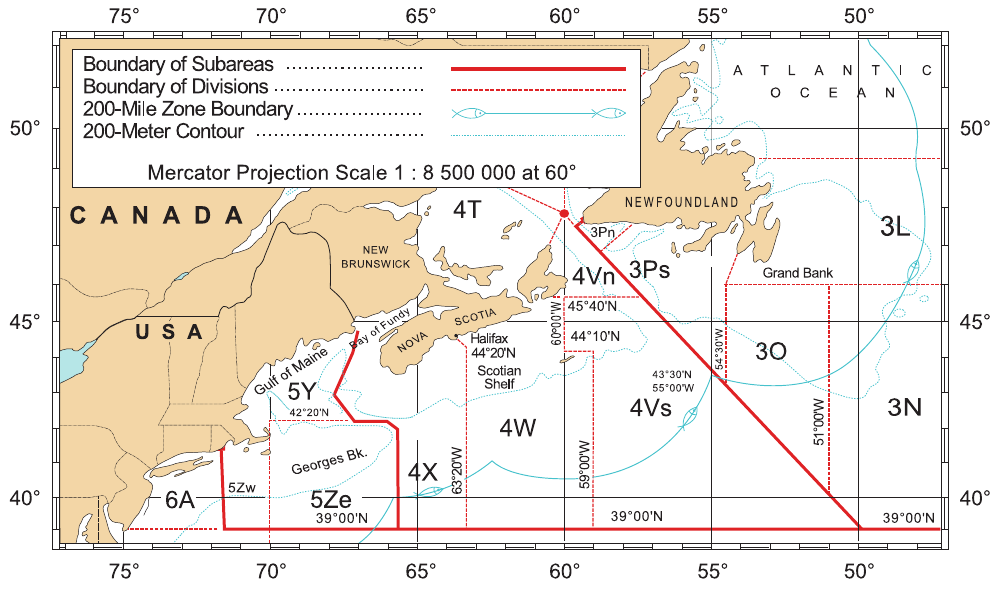}
	\caption[Map of the study region]{Map of the study region (\acs{NAFO} divisions 4Vn,s and 4W). Figure adapted from \url{http://www.nafo.int/about/frames/area.html}.}
	\label{fig:scotian_map}
\end{figure}

In order to evaluate SPRAT, we apply our model to explore observed fish stock dynamics on the eastern Scotian Shelf (\autoref{fig:scotian_map}) in the time period from 1970 to 2010. 

\begin{figure}[p]
	\centering
	\includegraphics[width=0.49\linewidth]{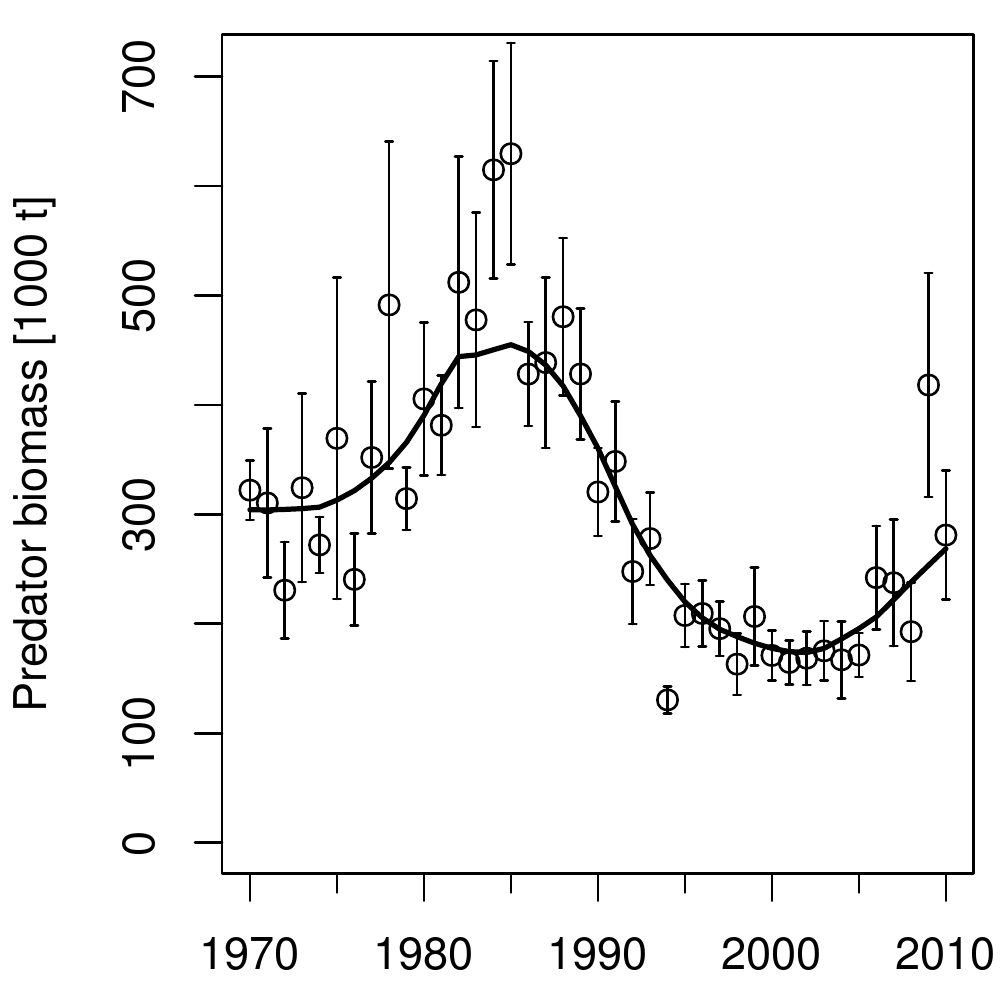}
	\includegraphics[width=0.49\linewidth]{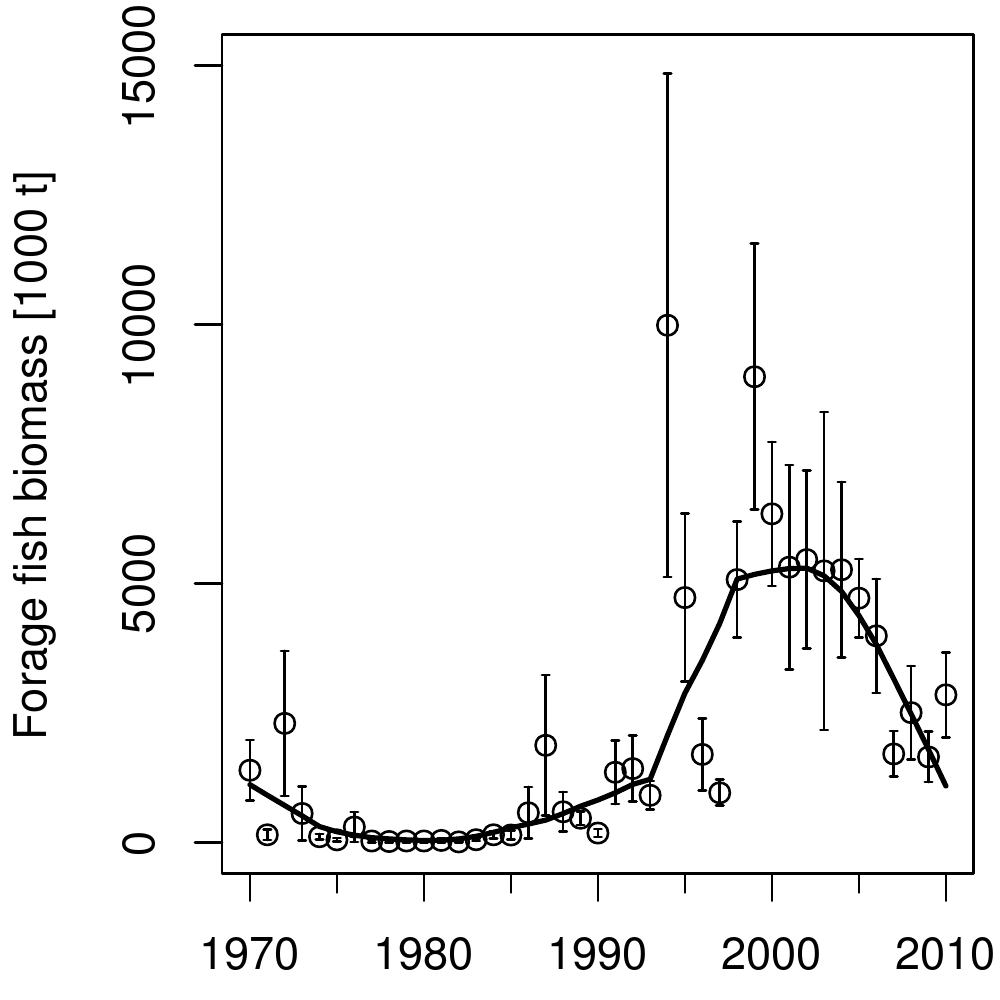}
	\caption[Observed biomass of demersal predators and forage fish]{Observed biomass of demersal predator and forage fish complexes as reported by \citet{frank2011}. The bold lines result from applying a 25~\% \acs{LOWESS} filter.}
	\label{fig:scotian_biomassObserved}
\end{figure}

\begin{figure}[p]
	\centering
	\includegraphics[width=\linewidth]{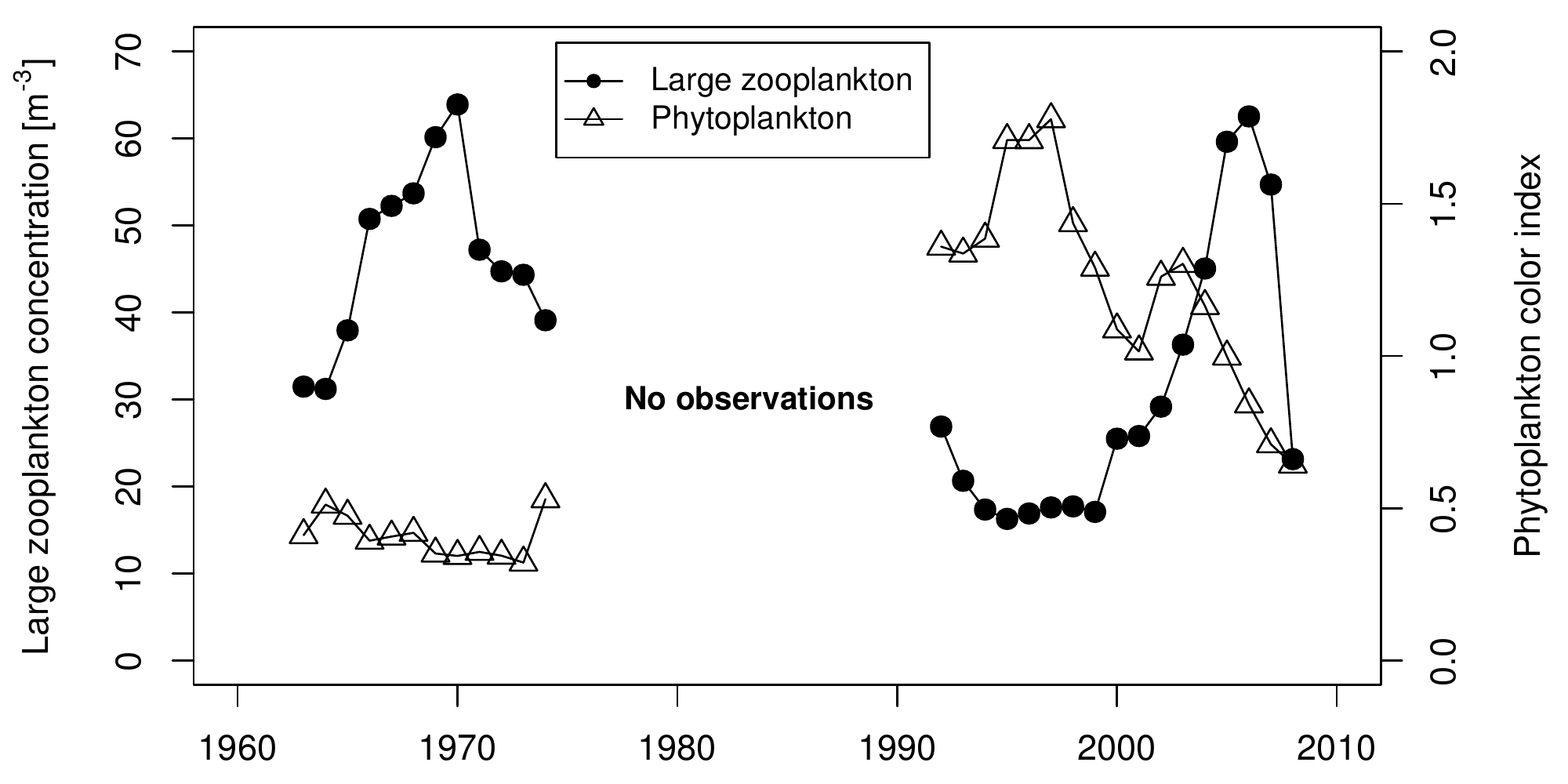}
	\caption{Observed large\hyp{}bodied zooplankton abundance and phytoplankton color index as reported by \citet{frank2011}.}
	\label{fig:scotian_observedPlankton}
\end{figure}

In the early 1990s, the eastern Scotian Shelf ecosystem underwent a distinct regime shift from a dominance of benthic predatory fish to a dominance of planktivorous forage fish species \citep{frank2011}. 
While benthic predators (predominantly Atlantic cod) declined, forage fish biomass (mainly herring \textit{Clupea harengus}) increased by up to 900~\% between the year 1990 and 2000 (see \autoref{fig:scotian_biomassObserved}). 
\citet{frank2005} found evidence that this regime shift was associated with a trophic cascade affecting not only piscivorous and planktivorous fish but also zooplankton and phytoplankton at the base of the food web. 
Due to reduced predation from the benthic predator fish complex, the forage fish complex could thrive, increasing its zooplankton consumption, which in turn reduced phytoplankton mortality associated with zooplankton grazing. 
As can be seen from \autoref{fig:scotian_observedPlankton}, large\hyp{}bodied zooplankton declined during the 1990s, while phytoplankton concentration increased at the same time. 

\begin{figure}[p]
	\centering
	\includegraphics[width=\linewidth]{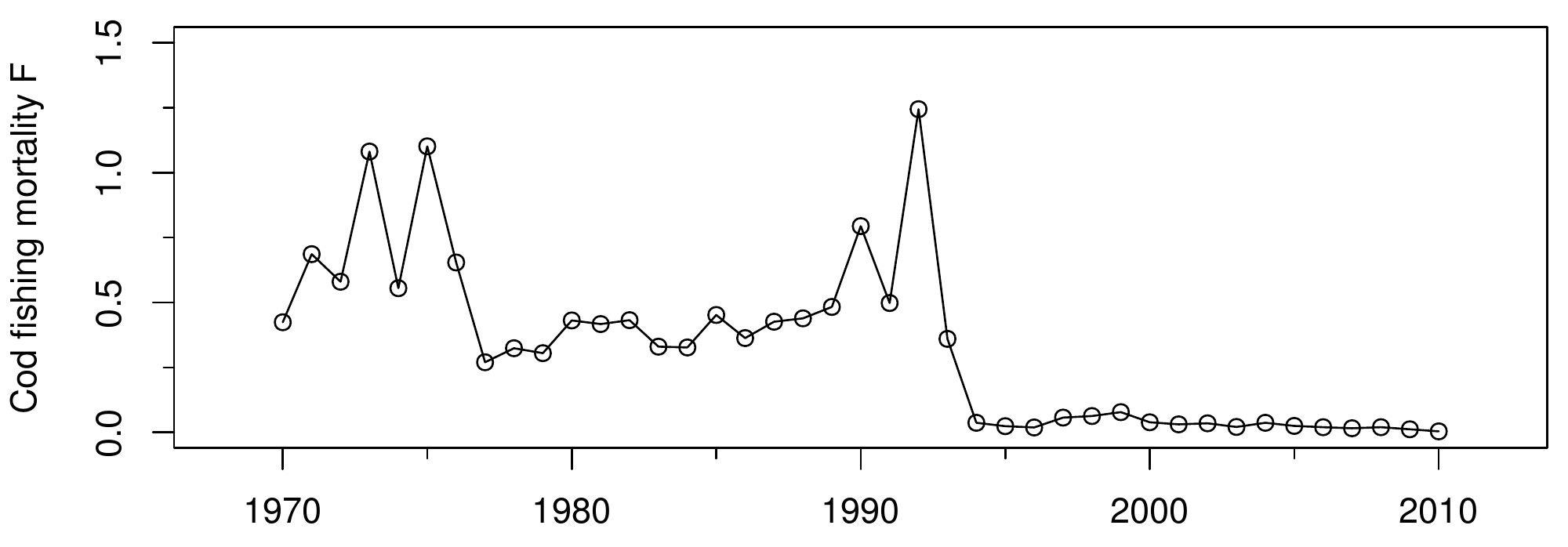}
	\caption{Cod fishing mortality as reported by \citet{frank2011}.}
	\label{fig:scotian_fishingMortality}
\end{figure}

Even though a fishing moratorium for cod and haddock was implemented in 1993 (see cod fishing mortality in \autoref{fig:scotian_fishingMortality}), the benthic predator stock complex did not show signs of recovery for over ten years. 
Only from 2006 on, the regime shift seems to have started reversing slowly across all four levels of the trophic cascade. 

\begin{figure}[p]
	\centering
	\includegraphics[width=\linewidth]{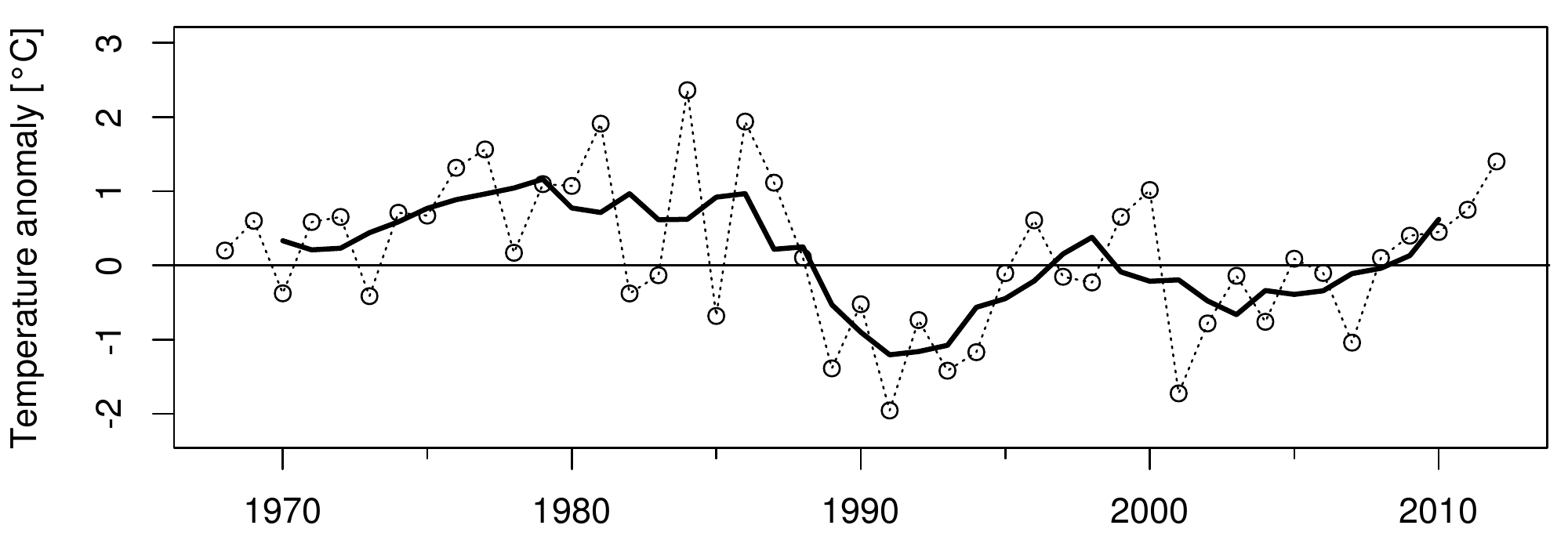}
	\caption{Observed bottom water temperature anomaly. The bold line is a five-year running mean. For details on how the data was derived, see \autoref{sec:parametrization}.}
	\label{fig:scotian_tempAnomaly}
\end{figure}
The occurrence of the shift itself is explained by most observers, such as \citet{frank2011}, as being caused by the intense fishing pressure on predator stocks prior to their collapse. 
Their findings suggest that environmental influences, such as the cooling of bottom waters (see \autoref{fig:scotian_tempAnomaly}), play only a minor role in driving the observed changes. 
While they recognize that the cooling is likely to have had a pronounced negative effect on benthic predator recruitment, their analysis indicates that this effect is dwarfed by the impacts of fishing. 

The prolonged duration of the collapse of the predator fish complex despite the closure of fisheries has been primarily attributed to predator\hyp{}prey reversal \citep{collie2013,minto2012}. 
When the predator stocks declined, forage fish were released from predation and their biomass level increased rapidly. 
Abundant forage fish are thought to have then competed with or directly preyed upon the early life stages of predator stocks, depressing their recruitment. 

An attempt at creating a model that can capture the fish stock dynamics described above was undertaken by \citet{bundy2005}. 
She employed a mass\hyp{}balance modeling approach using Ecopath with Ecosim \citep{christensen2005} to examine the trophic structure of the ecosystem prior and subsequent to the regime shift on the Scotian Shelf. 
Her model was able to reproduce the changes in the ecosystem structure during the regime shift but it offered little guidance for mechanistically \emph{explaining} the main drivers that lead to the shift. 
Therefore, our focus lies on applying the SPRAT model to investigate hypothesized drivers for the prolonged collapse and slow recovery of the benthic predator stocks on the Scotian Shelf. 

\subsection{Model Parametrization} \label{sec:parametrization}

We use SPRAT to simulate the fish stock dynamics on the eastern Scotian Shelf from 1970 to 2010 with two fish species. 
One of these species represents the benthic predator complex and the other the forage fish complex. 
Initially, the fish in the model are distributed evenly in the space\hyp{}size domain to achieve the biomass levels observed in 1970. 

\begin{figure}
	\centering
	\includegraphics[width=\linewidth]{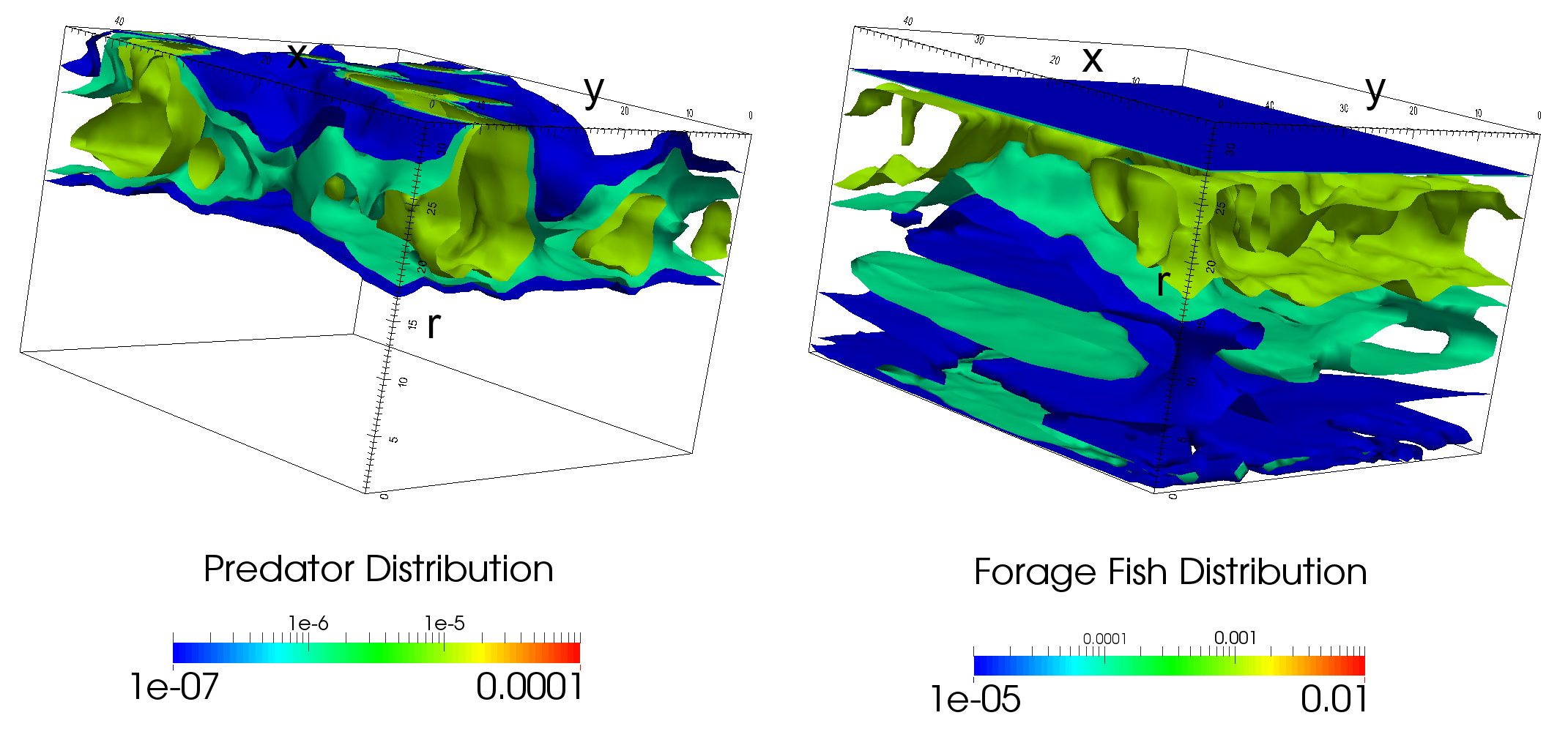}
	\caption{Visualization of predator and forage fish carbon biomass distributions from model output using isosurfaces. Note that for clarity all mesh cells are depicted to be of equal size in this figure. The $r$\hyp{}axis represents fish size and the unit of the scales is kg C m$^{-2}$ (kg C)$^{-1}$ = m$^{-2}$.}
	\label{fig:scotian_3DOutput}
\end{figure}

Space is assumed to be a homogeneous $450$ by $450$~km square, which roughly corresponds to the size of the continental shelf of \acs{NAFO} divisions 4Vn,s and 4W (see \autoref{fig:scotian_map}). 
The spatial domain is equipped with periodic boundary conditions and is discretized into $48$ by $48$ equally\hyp{}sized rectangular cells. 
The size dimension is divided into $32$ cells using logarithmically distributed division points. 
On this regular mesh, we employ piecewise linear finite elements (P1 elements) to approximate the solution of the model with a mass\hyp{}preserving \acf{FCT} \acf{FEM} solver \citep{brenner2008,kuzmin2012}. 
We determined the resolution of the mesh to be sufficient by running a series of control simulations with increasingly finer discretizations. 
The refinement process was stopped once the difference in simulated aggregated stock biomasses ceased to change noticeably. 
For a visualization of the three\hyp{}dimensional output produced by our model, see \autoref{fig:scotian_3DOutput}. 

For the simulations we report on in this paper, SPRAT was not coupled with a fully\hyp{}developed biogeochemical ocean model but instead used a simple \acs{NPZ} model (summarized in \appref{sec:appendixNPZ}) to represent the lower trophic levels of the ecosystem. 
Since we neglect currents in our simulations, the state variables of the \acs{NPZ} model do not have to be transported in space. 
Instead, we assign an instance of the \ac{NPZ} model to every discrete spatial point of the mesh for approximating the solution of the SPRAT model and compute the trophic interactions for each of these points in isolation. 
Note that we use only one set of parameters for all instances of the \acs{NPZ} model. 

The first step in the parametrization of SPRAT focuses on the parameters of the \acs{NPZ} model. 
The aim is to achieve a periodically stable evolution of the \acs{NPZ} state variables while the model is \emph{not} coupled with the fish model (a constant predation pressure by fish on $Z$ is assumed during parametrization of the NPZ model instead). 
Under these circumstances, the evolution of $N$, $P$, and $Z$ should match observations that can, for example, be obtained from the \acs{NOAA} World Ocean Database \citep{boyer2013}. 
All concentrations were integrated vertically, assuming the mixed ocean layer to be 40~m deep. 
To fit the model to these observations, we first extracted an average value for cloud coverage $C_c$ from the \acs{NASA} Earth Observations Database\footnote{\url{http://neo.sci.gsfc.nasa.gov/view.php?datasetId=MODAL2_M_CLD_FR}} for our model region. 
The other parameters of the \ac{NPZ} model were initially set to values obtained from \citet{franks2001} and were then manually adjusted to the values given in \autoref{tab:scotian_npzParameters} in \appref{sec:parameterTables} in order to match the observations from the \acs{NOAA} World Ocean Database. 
The \ac{NPZ} model proved to be most sensitive to the parameters associated with nutrient uptake ($V_m$, $k_s$) and with grazing of phytoplankton ($R_m$, $\lambda_s$).

In \autoref{tab:scotian_spratGlobalParameters} in \appref{sec:parameterTables}, we list the values of the global parameters of the fish model. 
The minimal predator\hyp{}prey mass ratio $\tau$ was estimated from FishBase \citep{froese2016}. 
The other two parameters, $r_{\tn{view}}$ and $\zeta_0$, were arbitrarily chosen by us since their values did not exhibit a strong influence on modeling results in our experiments. 
A radius of perfect information of $r_{\tn{view}} = 100$ km together with $\zeta_0 = 1$ kg$^{-1}$ leads to a balanced use of reactive and predictive movement strategies by the fish. 
The rather large radius of perfect information is a simplification that is justified by the fact that fish are able to learn information about their environment \citep[cf.\@][]{brown2008}.

Most species parameters of the fish model are directly observable and can be obtained from individual publications or from FishBase \citep{froese2016}, as seen in \autoref{tab:scotian_spratSpeciesParameters} of \appref{sec:parameterTables}. 
We used \textit{Gadus morhua} (Atlantic cod) and \textit{Clupea harengus} (Atlantic herring) to represent the predator and forage fish complexes, respectively. 
Some parameters, however, were utilized to manually fit the model to the observed fish biomasses described in \autoref{sec:scenario}. 
These parameters, which are mostly related to foraging, are marked with an asterisk in \autoref{tab:scotian_spratSpeciesParameters}. 

Some species parameters, such as the cruise speed or the beginning and end of the mating season, can only be measured with some uncertainty. 
To investigate how sensitive the model is to slight variations in these parameters, we ran several test simulations in which we altered the values of each of these variables in isolation and examined the effect on model output. 
For cruise speed $\varsigma$, slight variations (up to $\pm$20~\%) had only negligible influence on model results. 
The same is true for the mating season: 
as long as the mating seasons of the two fish complexes do not overlap, their placement throughout the year does not have a large influence on model results. 

For the fishing mortality of the predator fish complex, we apply a linear interpolation of the observed fishing mortality reported by \citet{frank2011} as shown in \autoref{fig:scotian_fishingMortality}. 
We assume zero fishing mortality for the forage fish complex. 

We found some disagreement in the literature between different published bottom water temperature anomaly series (especially between \citet{frank2011} and \citet{hebert2014} as well as \citet{zwanenburg2002}). 
Therefore, we reanalyzed temperature data from the Canadian Department of Fisheries and Oceans Oceanographic Database\footnote{\url{http://www.bio.gc.ca/science/data-donnees/base/index-en.php}} to obtain the most comprehensive and up\hyp{}to\hyp{}date time series for bottom water temperature anomaly $\Delta T(t)$. 
We filtered the raw data from the Oceanographic Database to acquire only temperature measurements that have been taken in the study area on the continental shelf east of Halifax, Nova Scotia in depths between 150~m and 250~m. 
To obtain yearly mean temperatures for this subset of the data, we fitted an \acf{ANOVA} model with temperature as the response variable and year and month as the predictor variables to the data \citep[for the modeling approach, cf.\@][]{worm2003}. 
The resulting yearly mean temperatures minus the overall mean temperature are plotted as open circles in \autoref{fig:scotian_tempAnomaly}. 
$\Delta T(t)$ itself is defined as the linear interpolation of the five\hyp{}year running means of these yearly temperature anomalies (this corresponds to the bold graph in \autoref{fig:scotian_tempAnomaly}). 
The resulting temperature anomaly is in agreement with data from individual observation stations on the eastern Scotian Shelf as reported by \citet{hebert2014,zwanenburg2002}. 

\section{Results}

As described in \autoref{sec:scenario}, three potential main drivers of the regime shift on the Scotian Shelf and its prolonged duration have been identified in the literature: 
\begin{enumerate}
	\item Intense fishing of benthic predators
	\item Predator\hyp{}prey reversal
	\item Pronounced cooling of bottom waters
\end{enumerate}
With the SPRAT model, we explore the effect of each of these potential drivers by disabling the corresponding functionality in the model (\eg, by setting $\Delta T \equiv 0$) and comparing the resulting model output to a reference simulation.
This reference simulation contains the full model functionality and uses the parametrization given in \autoref{sec:parametrization}.  

\subsection{Reference Simulation} \label{sec:refSim}

\begin{figure}[t]
	\centering
	\includegraphics[width=\linewidth]{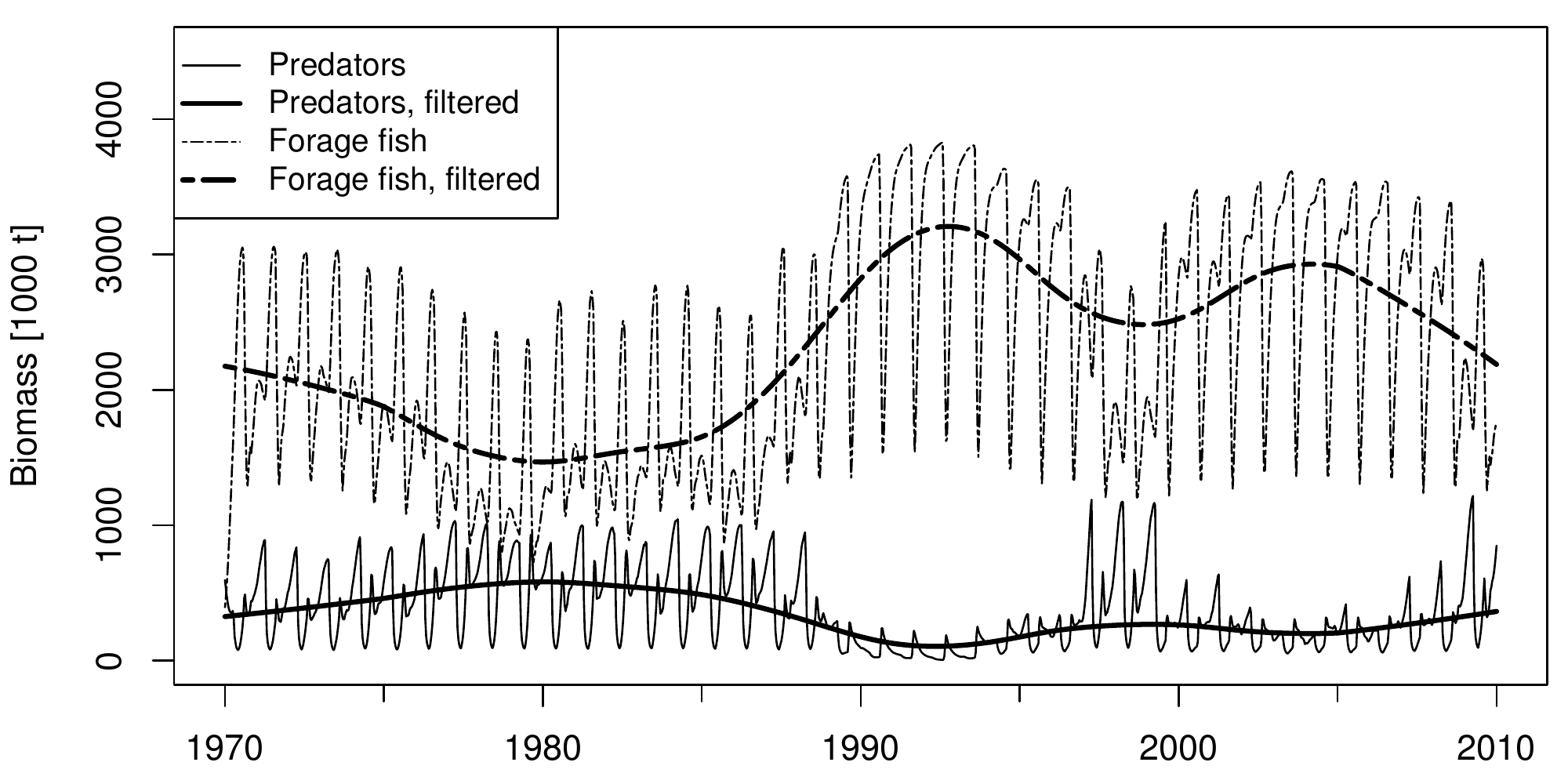}
	\caption[Aggregated biomass for the predator and the forage fish complex of our reference simulation]{Aggregated biomass for the predator and the forage fish complex of our reference simulation. The bold lines are the result of applying a 25~\% \acs{LOWESS} filter to the data.}
	\label{fig:scotian_biomassReferenceRaw}
\end{figure}

\begin{figure}[t]
	\centering
	\includegraphics[width=\linewidth]{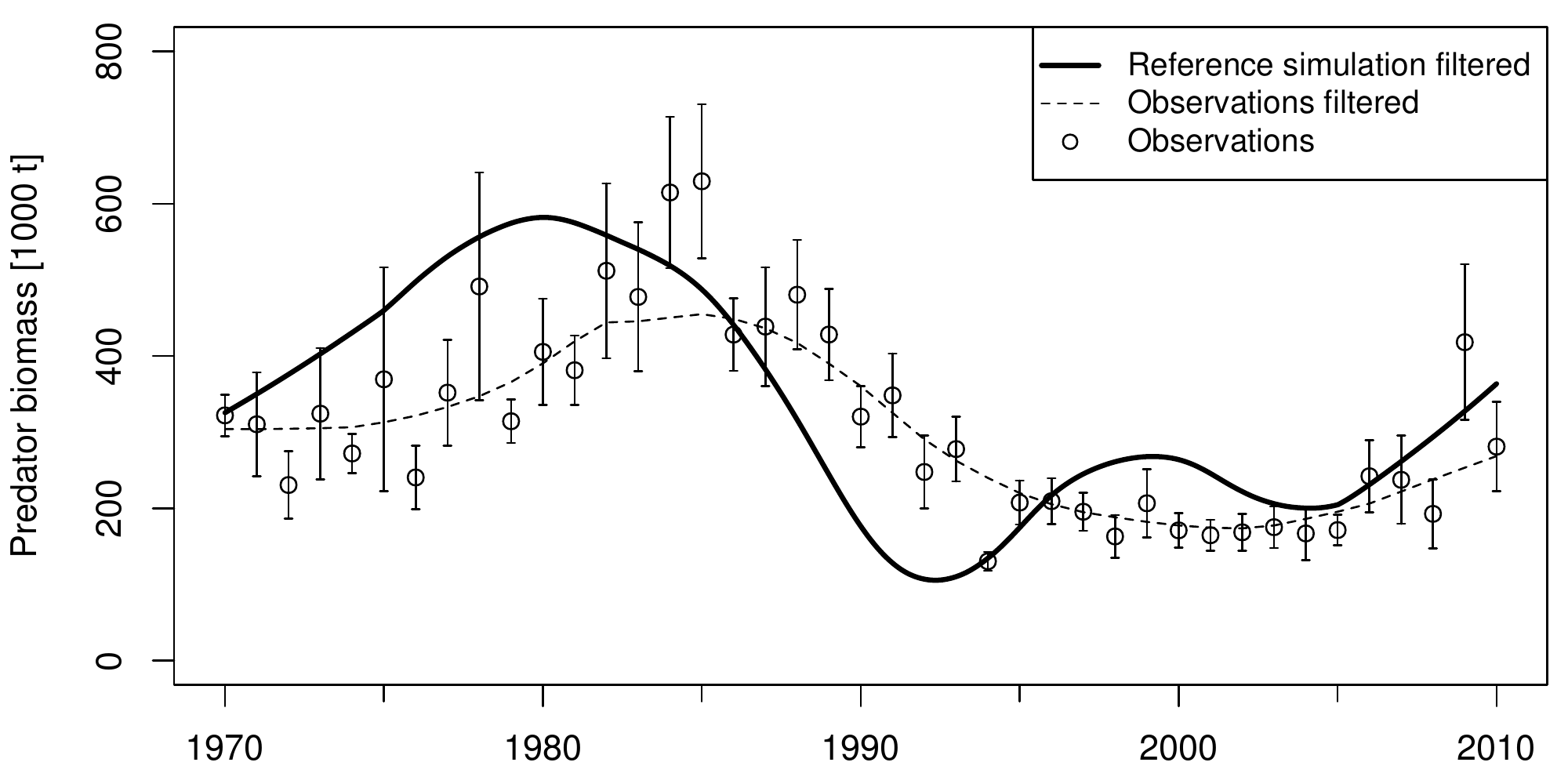}\\
	\includegraphics[width=\linewidth]{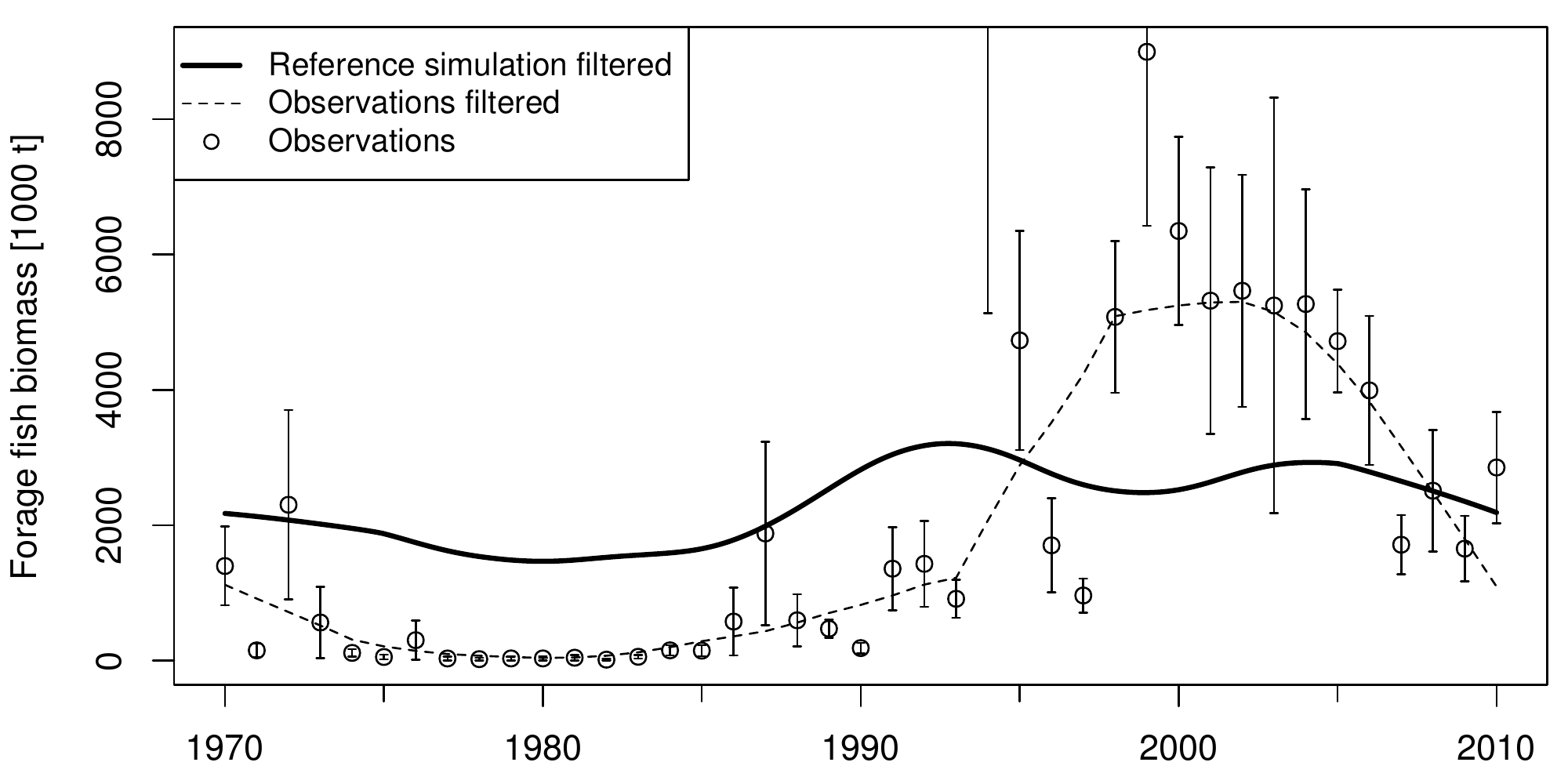}
	\caption[Filtered biomass time series from our reference simulation in comparison with observations]{Filtered biomass time series from our reference simulation in comparison with observations.}
	\label{fig:scotian_biomassReferenceSim}
\end{figure}

In \autoref{fig:scotian_biomassReferenceRaw}, we show the aggregated biomass of the predator and forage fish complexes of the reference simulation. 
Applying a 25~\% \acs{LOWESS} filter to our results removes seasonal variation in fish biomass and allows us to compare the results from our reference simulation with observational data in \autoref{fig:scotian_biomassReferenceSim}. 

Our model is able to qualitatively reproduce the shape of the observed biomass trajectories. 
However, predator biomass is overestimated prior to its collapse (although within the margin of error of the raw data) and underestimated at the beginning of the collapse. 
Additionally, the benthic predator biomass exhibits a small local maximum in the late 1990s that is not present in the Scotian Shelf observations but, interestingly, has been documented for nearby benthic predator stocks (\eg, for cod in the southern Gulf of St.~Lawrence as reported by \citet{swain2008}). 
The collapse of the benthic predators and the associated increase of forage fish biomass occurs a few years too early in our simulation. 
Forage fish biomass levels are significantly overestimated prior to the regime shift and underestimated during the shift (the model can only reproduce a doubling of forage fish biomass). 

\begin{figure}
	\centering
	\includegraphics[width=0.99\linewidth]{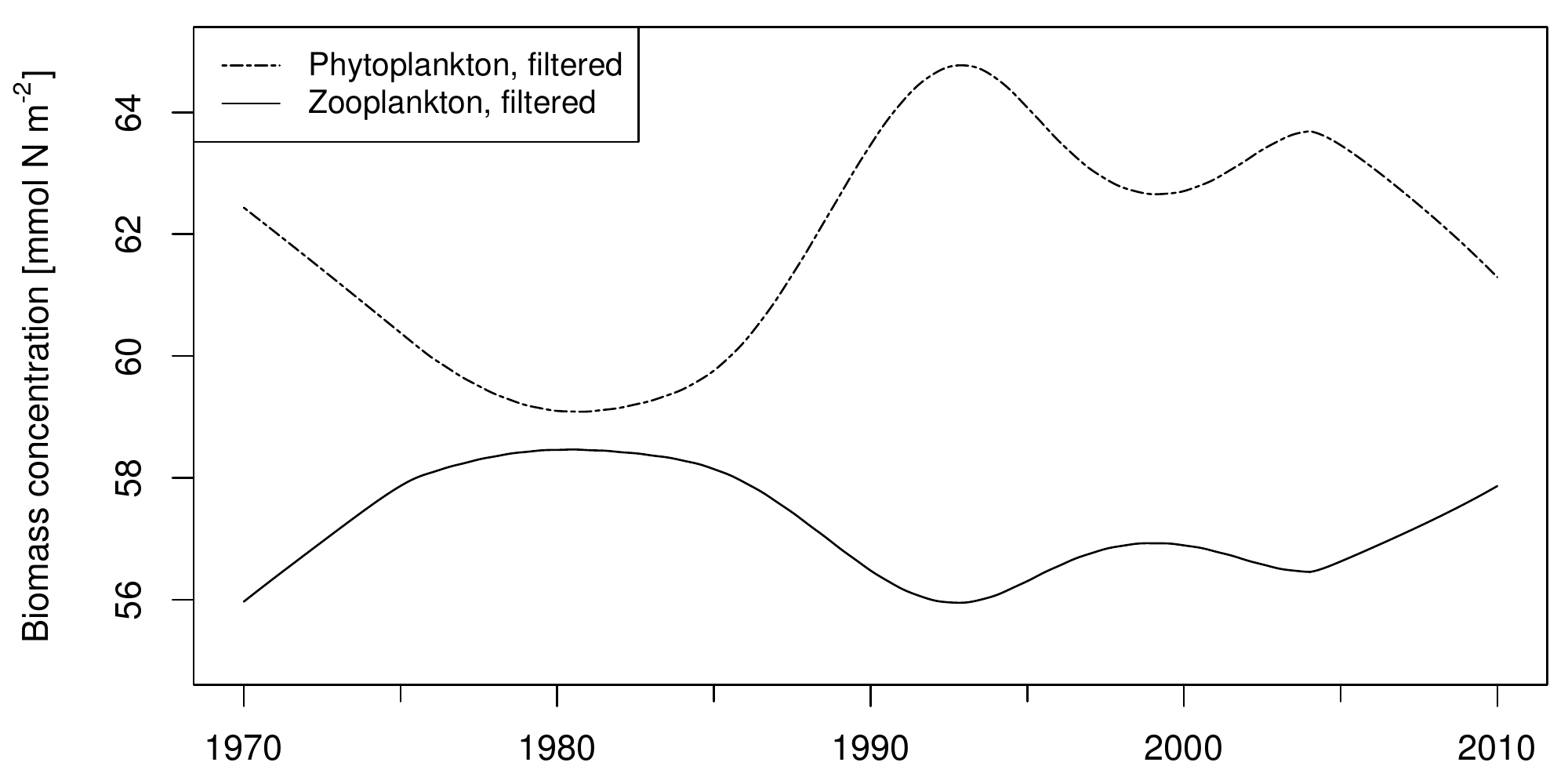}
	\includegraphics[width=0.99\linewidth]{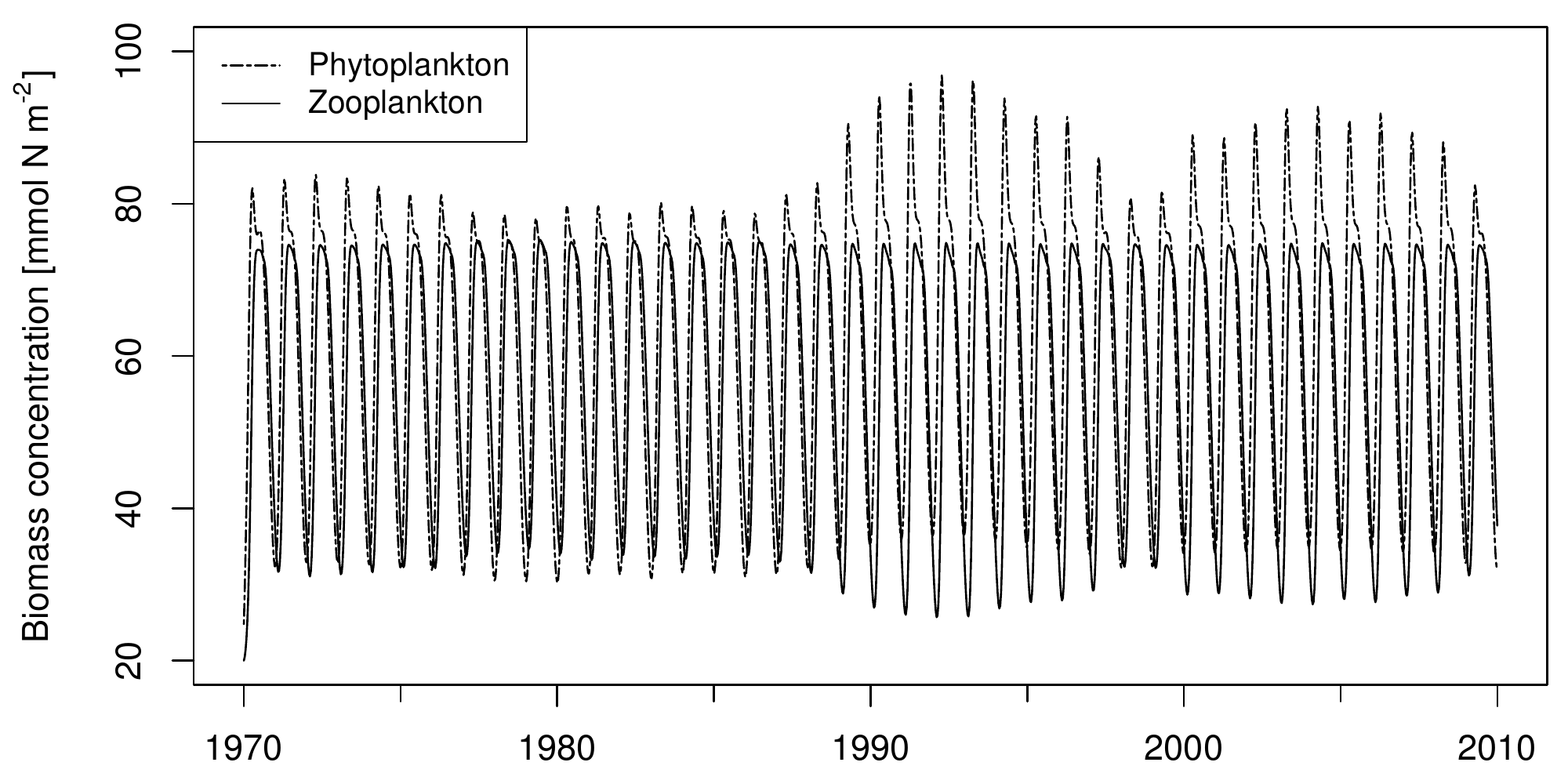}
	\caption[Aggregated zooplankton and phytoplankton abundance in our reference simulation]{Aggregated zooplankton and phytoplankton abundance in our reference simulation (top: 30~\% \acs{LOWESS} filtered; bottom: raw data).}
	\label{fig:scotian_simulatedPlankton}
\end{figure}

The observed shift in the plankton community from large\hyp{}bodied zooplankton to a more phytoplankton\hyp{}dominated regime is also present in the simulation data as can be seen in \autoref{fig:scotian_simulatedPlankton}. 
A quantitative comparison of our results with the observations shown in \autoref{fig:scotian_observedPlankton} is not possible because our \acs{NPZ} model does not resolve different zooplankton size classes. 

\subsection{Sensitivity Analysis} \label{subsec:sensitivityResults}

To determine which model parameters have the greatest influence on model results, we conducted a coarse sweep of the parameter space. 
Based on the reference simulation, we run simulations in which we halve and double the values of selected parameters (those used for fitting the model; see \autoref{tab:scotian_spratSpeciesParameters}).
For each of these simulations, we calculate the total misfit from fish biomass observations $\varepsilon_{\tn{obs}}$ via 
\begin{align}
	\varepsilon_{\tn{obs}} &= \sqrt{\left(\varepsilon_{\tn{obs}}^{[\tn{pred}]}\right)^2 + \left(\varepsilon_{\tn{obs}}^{[\tn{forage}]}\right)^2} &\text{with}\\
	\varepsilon_{\tn{obs}}^{[\tn{pred}]} &= \frac{\left\lVert B_i^{[\tn{pred}]} - B_{\tn{obs}}^{[\tn{pred}]} \right\rVert_{L_2, [0,t_{\max}]}}{\left\lVert B_{\tn{obs}}^{[\tn{pred}]} \right\rVert_{L_2, [0,t_{\max}]}} &\text{and} \\
	\varepsilon_{\tn{obs}}^{[\tn{forage}]} &= \frac{\left\lVert B_i^{[\tn{forage}]} - B_{\tn{obs}}^{[\tn{forage}]} \right\rVert_{L_2, [0,t_{\max}]}}{\left\lVert B_{\tn{obs}}^{[\tn{forage}]} \right\rVert_{L_2, [0,t_{\max}]}} \text{,} & 
\end{align}
where $B_{\tn{obs}}^{[\kappa]}$ are the \acs{LOWESS} filtered observed fish biomasses from \autoref{fig:scotian_biomassObserved}. 

\begin{figure}
	\centering
	\includegraphics[width=\linewidth]{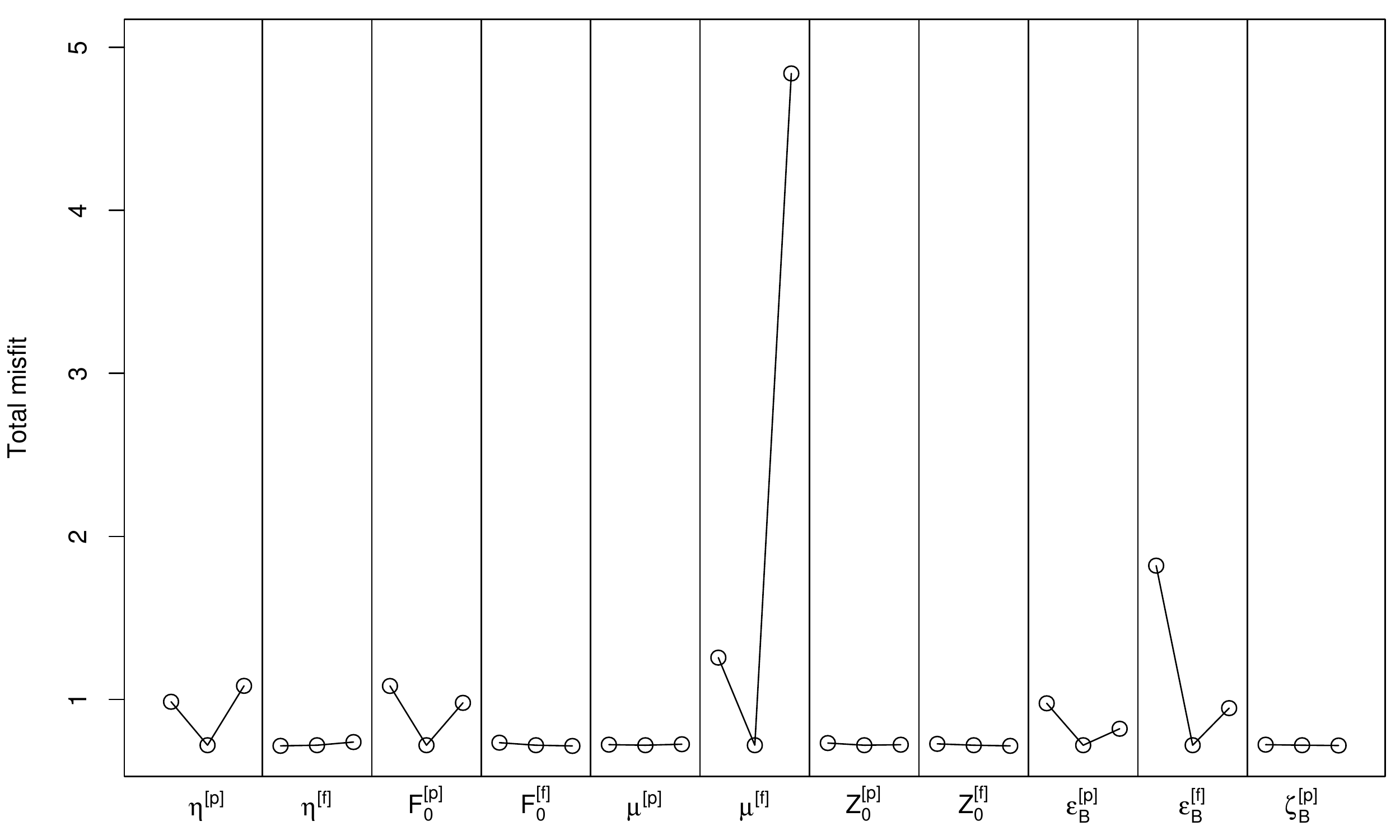}
	\caption[Sensitivity of the total model misfit to changes in several parameters]{Sensitivity of the total model misfit for the fish biomass time series to changes in several parameters. For each group of three values, the one in the center represents the reference simulation, the left one a simulation in which the respective parameter value is halved, and the right one a simulation in which the parameter value is doubled. The exponent $[\tn{p}]$ is for the predator complex and $[\tn{f}]$ is for the forage fish complex.}
	\label{fig:scotian_sensitivity}
\end{figure}
In \autoref{fig:scotian_sensitivity}, we plot $\varepsilon_{\tn{obs}}$ for the simulations grouped by model parameter. 
The steeper the lines connecting the misfit values are, the more sensitive the model is to the respective parameter. 
While some model configurations exhibit a smaller error with respect to observations than our reference simulation, we did not select them as our reference because they fail to capture all qualitative features of the observed biomass curves. 

Of the eleven parameters we ran sensitivity experiments for, the SPRAT model is most sensitive to $\mu^{[\tn{f}]}$, which controls the rate at which the forage fish are feeding on zooplankton. 
It is followed by the parameters for the background mortality rate $\varepsilon_B^{[\tn{f}]}$ and $\varepsilon_B^{[\tn{p}]}$, to which the model is second and fifth most sensitive, respectively. 
In between these two parameters, on the third and fourth place, lie $\eta^{[\tn{p}]}$ and $F_0^{[\tn{p}]}$, which govern the predation process of the predator complex. 
To the remaining parameters that were tested, the model is comparably insensitive. 

\subsection{Counterfactual Simulations}

In order to compare the effect size of the different hypothesized drivers of the regime shift on the Scotian Shelf, we report $L_2$ norm errors relative to the predator complex biomass of the reference simulation. 
For the aggregated predator complex biomass of the $i$-th simulation
\begin{equation}
	B_i^{[\tn{pred}]} (t) = \int_\Omega m^{[\tn{pred}]} (t,x,y,r) \; d(x,y,r)
\end{equation}
and that of the reference simulation $B_r^{[\tn{pred}]} (t)$, the relative $L_2$ error is given by 
\begin{equation}
	\varepsilon_{L_2} \left(B_i^{[\tn{pred}]}\right) = \frac{\left\lVert B_i^{[\tn{pred}]} - B_r^{[\tn{pred}]} \right\rVert_{L_2, [0,t_{\max}]}}{\left\lVert B_r^{[\tn{pred}]} \right\rVert_{L_2, [0,t_{\max}]}} \text{.}
\end{equation}

\subsubsection{The Influence of Temperature}

\begin{table}
	\centering
	\begin{tabular}{lrr}
		\toprule
		Scenario                    & $\varepsilon_{L_2}$ & $\frac{\varepsilon_{L_2}}{\varepsilon_{L_2} (\Delta T \equiv 0)}$ \\ \midrule
		$\Delta T \equiv 0$         &             $0.478$ &                                                               $1$ \\
		$H_{\tn{fish}} \equiv 0$ &             $0.228$ &                                                           $0.477$ \\
		No pred.\hyp{}prey reversal &             $0.183$ &                                                           $0.383$ \\ \bottomrule
	\end{tabular}
	\caption[Relative $L_2$ errors for the predator complex biomass]{Relative $L_2$ errors for the predator complex biomass and relative influence of the main drivers of the regime shift.}
	\label{tab:scotian_L2Errors}
\end{table}

\begin{figure}
	\centering
	\begin{subfigure}[b]{\linewidth}
	    \includegraphics[width=\linewidth]{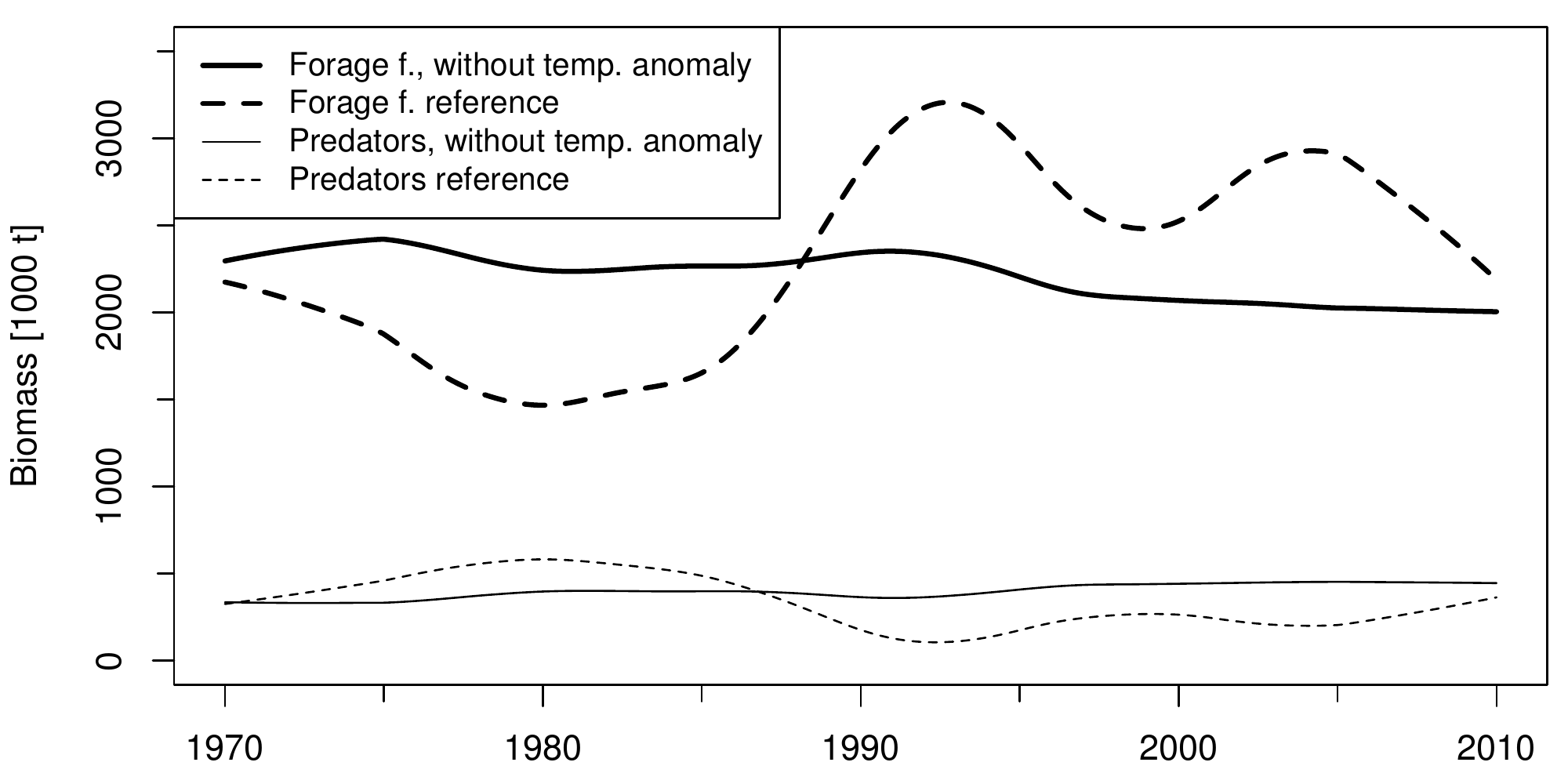}
    	\caption{Simulation with $\Delta T \equiv 0$.\vspace*{4pt}}
    	\label{fig:scotian_noTAnom}
	\end{subfigure}
	\begin{subfigure}[b]{\linewidth}
		\includegraphics[width=\linewidth]{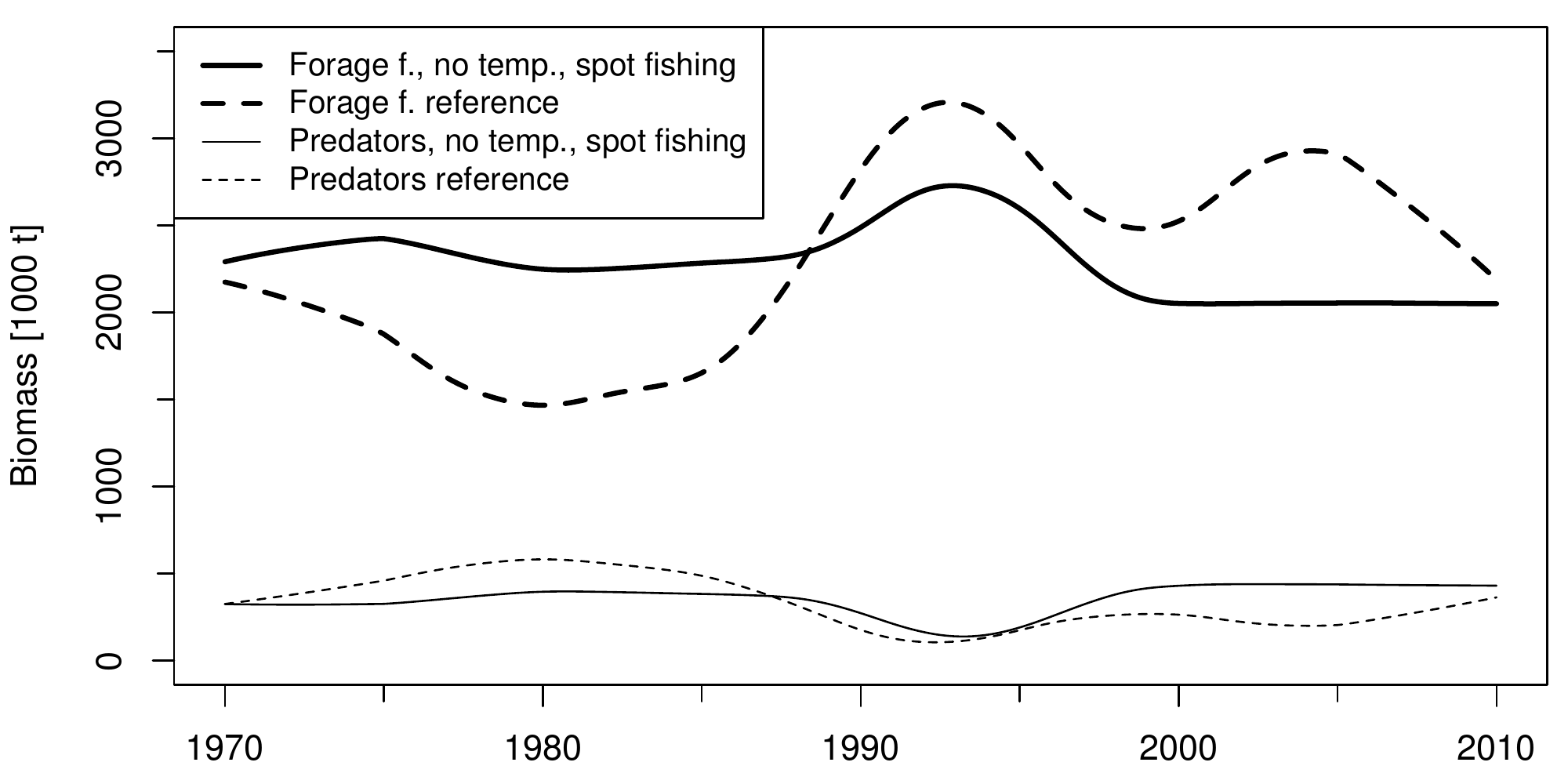}
		\caption{Simulation with $\Delta T \equiv 0$ and $F=20$ in 1992\vspace*{4pt}}
		\label{fig:scotian_noTAnomHighF}
	\end{subfigure}
	\begin{subfigure}[b]{\linewidth}
		\includegraphics[width=\linewidth]{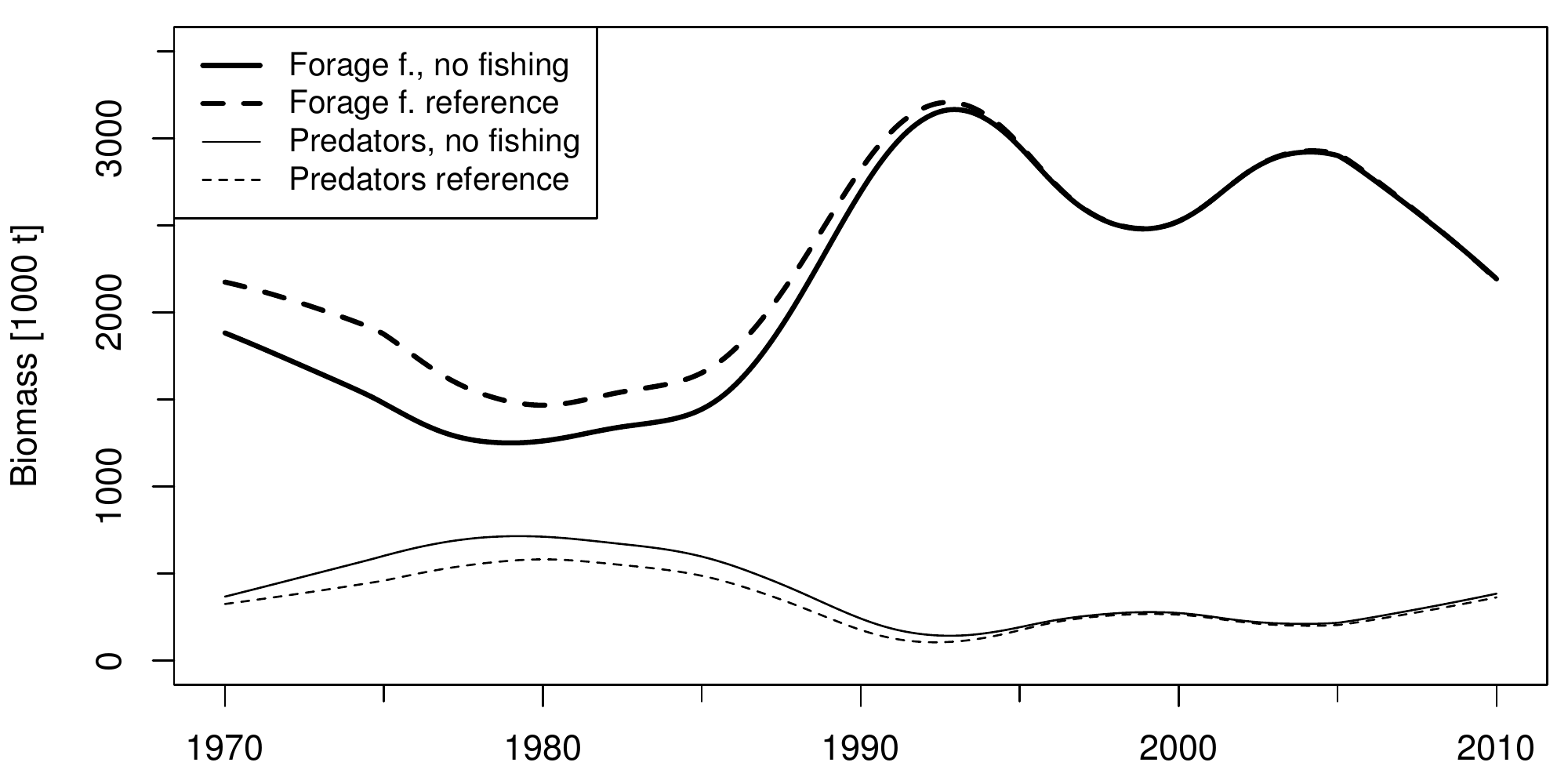}
		\caption{Simulation with $H_{\tn{fish}} \equiv 0$.\vspace*{4pt}}
		\label{fig:scotian_noFishing}
	\end{subfigure}
	\begin{subfigure}[b]{\linewidth}
		\includegraphics[width=\linewidth]{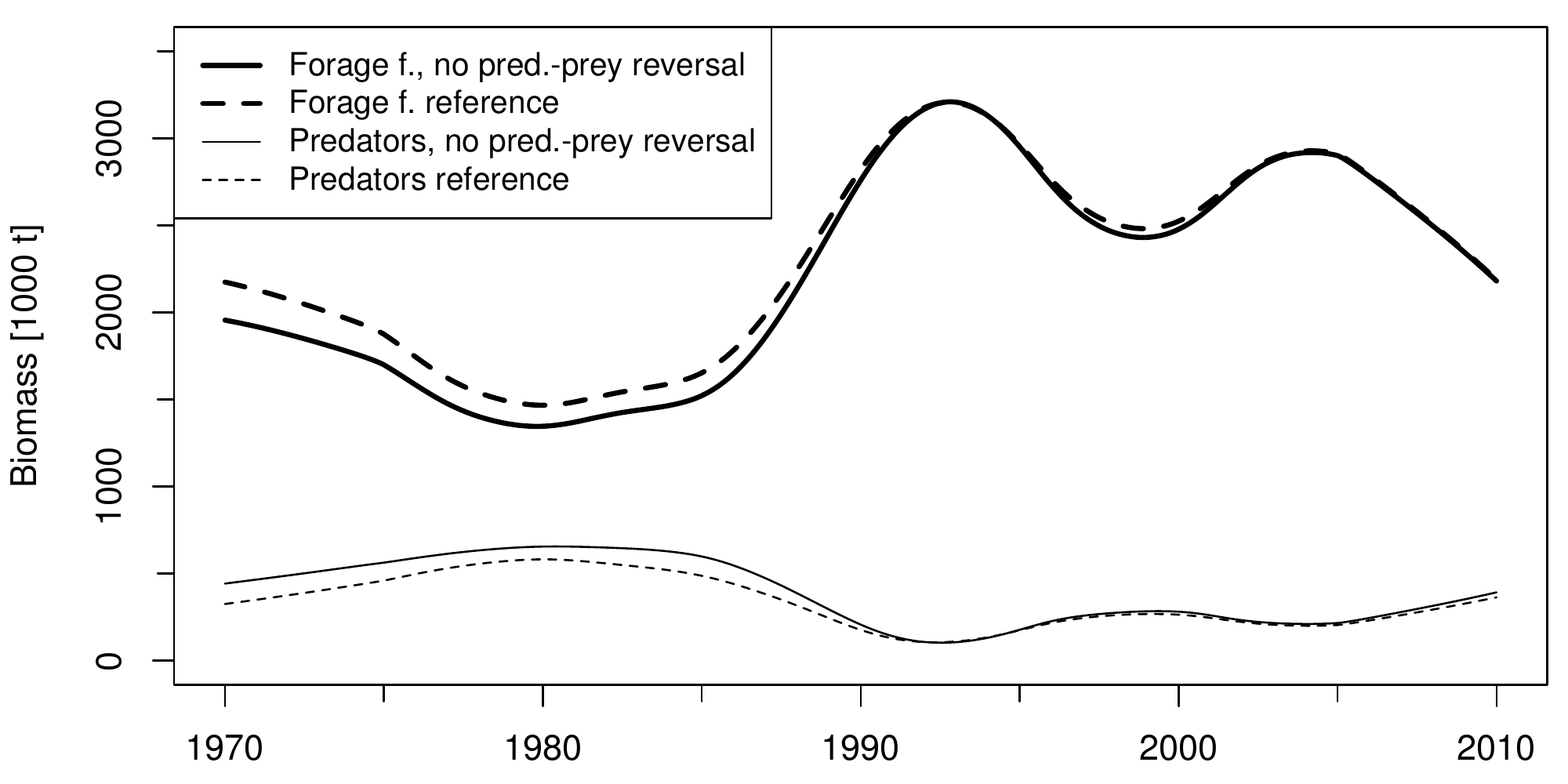}
		\caption{Simulation without predator\hyp{}prey reversal.}
		\label{fig:scotian_herringDoesNotEatLarvae}
	\end{subfigure}
	\caption{Different simulations in comparison with the reference simulation.}
	\label{fig:scotian_counterfactSims}
\end{figure}

To analyze the influence of the bottom water temperature anomaly as a driver of the regime shift, we set $\Delta T \equiv 0$. 
This results in the biomass curves depicted in \autoref{fig:scotian_noTAnom} and a relative $L_2$ error of $\varepsilon_{L_2} = 0.478$ as can be seen from \autoref{tab:scotian_L2Errors}.  
If we neglect the influence of the bottom water temperature anomaly, the regime shift is mostly non\hyp{}present.

To explore whether we can induce the regime shift by fishing alone without considering the temperature anomaly, we set the fishing intensity for benthic predators to $F=20$ in 1992 (see \autoref{fig:scotian_noTAnomHighF}).
However, even such an unrealistically high fishing pressure cannot introduce a persistent regime shift if we do not include the effects of bottom water temperature in our model. 

\subsubsection{The Influence of Fishing}

Setting $H_{\tn{fish}} \equiv 0$ in our model results in overall higher predator biomass levels and lower forage fish levels (see \autoref{fig:scotian_noFishing}). 
However, the difference compared to the reference simulation becomes neglectable after about 1993 (the year of the fishing moratorium) and the regime shift is still clearly expressed. 
The relative $L_2$ error for this simulation is $\varepsilon_{L_2} = 0.228$, which is less than half of that of the scenario with $\Delta T \equiv 0$ (see \autoref{tab:scotian_L2Errors}). 

\subsubsection{The Influence of Predator\hyp{}Prey Reversal}

To disable the last potential driver of the regime shift (predator\hyp{}prey reversal) in our model, we prevented forage fish from preying on predator juveniles with a wet mass of less than 120~g. 
As is apparent from \autoref{fig:scotian_herringDoesNotEatLarvae}, this leads to a situation similar to the non\hyp{}fishing scenario discussed above. 
The relative $L_2$ error of $\varepsilon_{L_2} = 0.183$ for the exclusion of predator\hyp{}prey reversal is even smaller than for the no\hyp{}fishing scenario and is only about 38~\% of the relative $L_2$ error of the $\Delta T \equiv 0$ scenario. 

\section{Discussion}

The observed dynamics of the eastern Scotian Shelf ecosystem are markedly complex, involving a rapid and lasting regime shift, a possible predator\hyp{}prey reversal and non\hyp{}recovery of a formerly dominant species. 
Given this complexity, the reasonable fit of our reference simulation results (\autoref{sec:refSim}) with observed biomass trends may be viewed as more than satisfactory. 
The SPRAT model is able to reproduce all dynamic features that are present in the observational data, and the model fits the observed predator complex biomass data mostly within the margin of observational error. 
Since we are mainly interested in exploring possible causes of the regime shift, a more exact match of the observed forage fish biomass trajectory may not be fully relevant as long as its basic shape is reproduced, and provided that the sensitivities of the modeled processes are simulated correctly. 
Note, however, that we simplify the complex structure of the Scotian Shelf ecosystem by taking into account only two fish species complexes, which places considerable constraints on the range of possible model outcomes and limits the predictive power of our model as discussed below.

\subsection{Drivers of the Fish Stock Dynamics} \label{subsec:scotian_discussionDrivers}

Our results indicate that modeling both species interactions and fishing, as well as environmental parameters, such as bottom water temperature anomaly, all play a important role in explaining the regime shift and the biomass fluctuations associated with it. 
Yet, their strength may vary considerably.
In our set of simulations, and under the parameter values chosen here, the influence of ambient temperature on the benthic predator biomass dynamics was stronger than the influence of fishing and predator\hyp{}prey reversal alone or in combination. 
If the observed bottom water temperature anomaly is ignored ($\Delta T \equiv 0$), the regime shift is not reproduced in our model. 
However, if either fishing or predator\hyp{}prey reversal is disabled, the regime shift is still visible, if weakened. 

Such an impact of temperature changes on biomass levels of cod is in broad agreement with empirical findings of \citet{pershing2015slow}, who show\sdash{}for the Gulf of Maine\edash{}that sudden and rapid temperature changes can negatively affect cod recruitment success and increase mortality (observe however, that these events are not strictly comparable because the Gulf of Maine experienced a warming, the Scotian Shelf a cooling event). 
The comparably low impact of fishing on the biomass levels in our model, however, is in disagreement with the general observation that fish stocks react quite sensitively to changes in exploitation rates \citep{jennings1998}. 
We therefore have to caution that the model likely overestimates the influence of temperature in comparison with fishing. 
A possible explanation for this is that $\Delta T$ directly affects the predation (and zooplankton grazing) rates (see \eqnsref{eq:spratModel_predTemp} and \ref{eq:spratModel_grazingTemp}), which are among the most sensitive model parameters (\autoref{subsec:sensitivityResults}). 
The sensitivity of the model to the predation rates can, in turn, be explained by the fact that the modeled biomass levels of the fish stocks decline quite rapidly during winter (\autoref{fig:scotian_biomassReferenceRaw}). 
In spring, the stocks have to regenerate by taking up the right amount of biomass. 
If predation and grazing rates are even only marginally too low, the stocks cannot regenerate (leading to a large model misfit). 
If predation and grazing rates are marginally too high, stock biomass overshoots (again, leading to a large model misfit). 
The rapid decline of modeled fish biomass in winter can be attributed to low levels of primary productivity (\autoref{fig:scotian_simulatedPlankton}), which in effect deprive the fish of their food sources. 
In reality, however, stock biomass would not decrease that strongly because the fish might migrate further away from the shelf region to find subsidiary food sources \citep{sinclair1985}. 
In our model, however, we (wrongly) assume a homogeneous space with no possibility of migrating out of the study area. 

In conclusion, the dominant influence of temperature in comparison with fishing in our model can be traced back to a trade\hyp{}off between model complexity (ignoring migration) and accuracy. 
In order to correct the relative influence of temperature in the model, one would either have to parametrize migration from and towards the study area or explicitly include areas with subsidiary food sources for the winter months into the modeled region. 

Regardless of the possibly overemphasized role of $\Delta T$, our modeling results support the hypothesis that the pronounced cooling of bottom waters may have played some role in the regime shift on the Scotian Shelf, most likely in combination with elevated fishing pressure, such as observed two decades later on the Gulf of Maine \citep{pershing2015slow}. 
More generally, we might hypothesize that a complete restructuring of a marine ecosystem might often be related to multiple causalities that act in synergy. 
If this hypothesis held true, fishing might play a key role in decimating stocks but these stocks would have to be weakened by other factors to induce a long\hyp{}lasting collapse and associated long\hyp{}term ecosystem effects. 

Much of this interpretation, however, hinges on the particular temperature data that are used. 
As discussed in \autoref{sec:parametrization}, we recalculated the bottom water temperature anomaly $\Delta T$ from raw data because of conflicting temperature series reported in the literature. 
The results of our recalculation (\autoref{fig:scotian_tempAnomaly}) agree with data for individual observation stations on the eastern Scotian Shelf as documented by \citet{hebert2014,zwanenburg2002}. 
However, our results differ from the bottom water time series used by \citet{frank2011} in that we observe a more pronounced and longer\hyp{}lasting cooling effect in the early 1990s. 
If we force our model with the bottom water temperature data of \citet{frank2011}, the collapse of the predator fish complex is not reproduced in the model results. 
Therefore, the described differences between our temperature time series and the one by \citet{frank2011} could explain why we find a stronger correlation of the bottom water temperature with benthic predator biomass. 

\subsection{Sensitivity Analysis of the Model}

Above, we already discussed the sensitivity of SPRAT to predation and zooplankton grazing, specifically in the context of our analysis of the comparably large influence of $\Delta T$. 
In this context, we would like to add that the well\hyp{}defined maximum of zooplankton abundance (\autoref{fig:scotian_simulatedPlankton}) can likely be explained by the density\hyp{}dependent mortality included in our \ac{NPZ} model (see \autoref{eq:spratModel_dZ} in \appref{sec:appendixNPZ}). 
As the zooplankton abundance approaches its maximum, its mortality increases quadratically until it offsets any positive zooplankton increase. 

Regarding the results of the sensitivity experiments reported in \autoref{subsec:sensitivityResults} in general, we can observe that they seem to reflect the main trophic interactions one would expect to see in the eastern Scotian Shelf ecosystem. 
The main control mechanisms in the ecosystem are the bottom\hyp{}up forcing of the availability of zooplankton to the planktivorous forage fish and the top\hyp{}down control of predatory fish preying on the forage fish complex. 
Therefore, it seems plausible that, as we found in \autoref{subsec:sensitivityResults}, parameters associated with these processes ($\mu^{[\tn{f}]}$, $\eta^{[\tn{p}]}$, and $F_0^{[\tn{p}]}$) have a large effect on model results. 
We observe the strongest sensitivity for $\mu^{[\tn{f}]}$ because it is essentially the only parameter that regulates the import of zooplankton biomass into the fish model (therefore, increasing $\mu^{[\tn{f}]}$, dramatically increases overall fish biomass). 
The parameters $\mu^{[\tn{p}]}$, $Z_0^{[\tn{p}]}$, $\eta^{[\tn{f}]}$, and $F_0^{[\tn{f}]}$ exhibit only low sensitivity because the corresponding trophic interactions in the ecosystem are relatively weak. 
The only exception to this rule is $Z_0^{[\tn{f}]}$ (zooplankton grazing half saturation for forage fish), which belongs to a strong trophic link but to which the model is not sensitive. 
A possible explanation is that zooplankton levels are in saturation with respect to $Z_0^{[\tn{f}]}$ throughout most parts of the year (we choose $Z_0^{[\tn{f}]}$ relatively small) and that the same is still true for half and twice the value of the parameter. 

The high sensitivity of the model with respect to $\varepsilon_B^{[\tn{f}]}$ and $\varepsilon_B^{[\tn{p}]}$ is also related to their influence on (implicit) trophic links: 
they parametrize predation by birds and marine mammals, which is quite strong in the study area \citep{frank2011}. 

While we saw that the bottom water temperature anoma\-ly $\Delta T$ has a large impact on model results, the SPRAT model is insensitive to $\zeta_B^{[\tn{p}]}$\sdash{}the scaling constant of the $\Delta T$\hyp{}dependent predator juvenile mortality. 
This small impact of direct mortality due to $\Delta T$ supports our argument from above that the large overall effect size of the temperature anomaly is due to its influence on foraging processes.

\section{Conclusions}

With SPRAT, we introduced a spatially\hyp{}explicit fish stock model for end\hyp{}to\hyp{}end modeling that is based on \acfp{PBE}. 
As such, our model offers an alternative to the \acp{IBM} and \ac{ADR}\hyp{}based fish stock models commonly employed in this domain. 
By utilizing \acp{PBE}, SPRAT can rely on the advanced mathematical theory of \acp{PDE} with its good integration with existing biogeochemical models while still being formulated from the perspective of the individual fish with a dynamic food web structure and, thus, combines the advantages of \acp{IBM} and \ac{ADR} models. 

With regard to the application of SPRAT on the Scotian Shelf, we have shown that our model is able to reproduce the complex direct and indirect dynamics of the two major fish groups on the eastern Scotian Shelf, while capturing at least qualitatively a lasting regime shift that was observed there. 
Our results indicate that SPRAT is a promising tool for mechanistically exploring some of the processes that may be restructuring marine ecosystems and for exploring the possible effects of planned management interventions via ecosystem simulations. 
These efforts can, of course, only be a first step for the validation and improvement of our model. 
A promising but so far neglected direction for future research in this area is the empirical comparison of different fish stock models for end\hyp{}to\hyp{}end modeling by parametrizing the models for defined scenarios and comparing their outputs. 
This would allow us to evaluate how different design choices affect model predictions. 

Concerning the empirical understanding of the regime shift in the Scotian Shelf ecosystem, our model highlights the hypothesis that a pronounced cooling of bottom waters might have been a necessary condition for the prolonged collapse of the cod stocks by making them more vulnerable to the effects of fishing and the lasting impacts of a predator\hyp{}prey reversal between cod and herring, as well as other forage fish. 
Our results emphasize the importance of taking into full consideration the complex interplay of different environmental drivers and their cumulative impacts on marine resources as well as their supporting ecosystems \citep[cf.\@][]{britten2016changing,pershing2015slow}.

\appendix

\section{Summary of the NPZ Model for the Eastern Scotian Shelf} \label{sec:appendixNPZ}

Our \ac{NPZ} model contains only the three state variables given in \autoref{tab:spratModel_npzParam} and does not explicitly feature space. 
When coupling the \ac{NPZ} model with the spatially\hyp{}explicit SPRAT model, we assume\sdash{}for computational simplicity\edash{}that there are no currents in the ocean and that no diffusion takes place. 
Therefore, we can assign an instance of the \ac{NPZ} model to every discrete spatial point of the mesh for approximating the solution of the SPRAT model and compute the trophic interactions for each of these points in isolation. 
For converting between nitrogen concentrations and carbon mass concentrations, we use the conversion factor 
\begin{equation}
	\rho = 12 \cdot 10^{-6} \cdot \frac{106}{16} 
	\, \frac{\tn{kg C m}^{-2}}{\tn{mmol N m}^{-2}} \text{,}
\end{equation}
which is based on the Redfield stoichiometric ratio for plankton \citep{redfield1966}. 
\begin{table}
	\centering
	\begin{tabular}{lll}
		\toprule
		Symbol & Description   & Unit            \\
		\midrule
		$N(t)$ & Nutrients     & mmol N m$^{-2}$ \\
		$P(t)$ & Phytoplankton & mmol N m$^{-2}$ \\
		$Z(t)$ & Zooplankton   & mmol N m$^{-2}$ \\
		\bottomrule
	\end{tabular}
	\caption[State variables of the \acs{NPZ} model]{State variables of the \acs{NPZ} model.}
	\label{tab:spratModel_npzParam}
\end{table}

The general time\hyp{}dependent dynamics of the three state variables are given by 
\begin{align}
	\frac{dP}{dt} &= f(\iota)g(N)P - h(P)Z - i(P)P \\
	\frac{dZ}{dt} &= \gamma h(P)Z - j(Z)Z - \frac{1}{\rho} G \label{eq:spratModel_dZ} \\
	\frac{dN}{dt} &= -f(\iota)g(N)P + (1-\gamma) h(P)Z \\
		&\quad + i(P)P + j(Z)Z + r(N) \text{,}
\end{align}
where the meaning of the symbols is described in \autoref{tab:spratModel_npzSymbols}. 
\begin{paramTable}
	\begin{tabular}{lll}
		\toprule
		Symbol     & Description                          & Unit                     \\ 
		\midrule
		$\iota(t)$ & Daily integrated solar irradiance    & MJ m$^{-2}$              \\
		$f(\iota)$ & Phytoplankton response to irradiance & dimensionless            \\
		$g(N)$     & Phytoplankton nutrient uptake        & s$^{-1}$                 \\
		$h(P)$     & Zooplankton grazing rate             & s$^{-1}$                 \\
		$i(P)$     & Phytoplankton mortality rate         & s$^{-1}$                 \\
		$j(Z)$     & Zooplankton natural mortality rate   & s$^{-1}$                 \\
		$r(N)$     & Vertical nutrient transport rate     & mmol N m$^{-2}$ s$^{-1}$ \\
		$\gamma$   & Zooplankton assimilation efficiency  & dimensionless            \\
		$G(t)$     & Plankton consumption rate by fish    & mmol N m$^{-2}$ s$^{-1}$ \\ 
		\bottomrule
	\end{tabular}
	\caption[Functions and parameters of a general \acs{NPZ} model]{Functions and parameters of a general \acs{NPZ} model.}
	\label{tab:spratModel_npzSymbols}
\end{paramTable}
The transfer functions ($\, f$, $g$, $h$, $i$, $j$, $r$, and $G$) define the mass fluxes (expressed via nitrogen content) between the different compartments of the model. 

The effect of light on the growth process is modeled via the saturating response function 
\begin{equation}
f(\iota) = \frac{\iota}{\iota_s + \iota} \text{,}
\end{equation}
where $\iota_s$ is the half saturation constant for the daily integrated irradiance available for photosynthesis $\iota$. 
We model the daily integrated solar irradiance according to \citet{brock1981}. 
For calculating the amount of irradiance actually available for photosynthesis, we consider three attenuation factors:
\begin{enumerate}
	\item Transmission losses: even without clouds, only a fraction of the radiation is actually transmitted from the top of the atmosphere through the surface of the ocean \citep[we assume $75$~\% to be transmitted; cf.][]{evans1985}.
	\item \acf{PAR}: phytoplankton can utilize only certain parts of the transmitted wavelength spectrum for photosynthesis \citep[we assume $50$~\% of the wavelength spectrum to be suitable for photosynthesis; cf.][]{fasham1990}. 
	\item Cloud cover: clouds decrease the amount of radiation that reaches the ocean surface. For modeling this aspect, we follow \citet{reed1977}, who derives the empirical relation 
	\begin{equation}
		\frac{Q_C}{Q_0} = 1 - 0.62 \cdot C_c + \frac{0.342}{\pi} \cdot Z_1 \text{,}
	\end{equation}
	where $Q_C$ is the insolation under cloudy conditions, $Q_0$ the insolation under clear skies, $C_c$ the percentage of cloud coverage, and $Z_1$ the solar altitude at noon (in radians). 
\end{enumerate}

Uptake of nutrients by phytoplankton is given by the Michaelis\hyp{}Menten equation 
\begin{equation}
g(N) = \frac{V_m N}{k_s + N}
\end{equation}
with $V_m$ being the maximum uptake rate and $k_s$ the half saturation constant for the available nutrients. 

The rate of phytoplankton consumption by zooplankton is given via Ivlev's functional response 
\begin{equation}
h(P) = R_m (1 - \exp(-\lambda_s P)) \text{,}
\end{equation}
where $R_m$ is the maximum phytoplankton uptake rate and $\lambda_s$\sdash{}the Ivlev constant\edash{}influences how fast saturation is achieved. 
The actual fraction of phytoplankton that is incorporated into the zooplankton biomass is influenced by the assimilation efficiency $\gamma$. 
The remaining part ($(1-\gamma) h(P)Z$) is excreted and, therefore, added to the nutrient pool again. 

Phytoplankton mortality is modeled via the quadratic density\hyp{}dependent mortality rate 
\begin{equation}
i(P) = \varepsilon_P P \text{.}
\end{equation}
For zooplankton, we define two mortality rates: 
first, a quadratic rate 
\begin{equation}
j(Z) = \varepsilon_Z Z
\end{equation}
that is directly applied in the model in \autoref{eq:spratModel_dZ}. 
Second, a linear rate 
\begin{equation} \label{eq:spatModel_ZDeathFish}
j_{f}(Z) = \varepsilon_f
\end{equation}
that is used to limit the amount of zooplankton consumed by fish $G$. 

For reasons of simplicity and model stability, we do not explicitly describe the vertical transport of water and remineralization processes associated with this transport but parametrize them via 
\begin{equation}
r(N) = \eta_r (N_m - N) \text{,}
\end{equation}
where $\eta_r$ sets the speed of remineralization and $N_m$ is the maximum nutrient concentration. 
The functional form of $r(N)$ mimics diffusion processes by assuming a constant nutrient concentration $N_m$ outside the region covered by the \acs{NPZ} model (\eg, in the deep ocean). 
Depending on whether $N$ is below or above $N_m$ in the model, nutrients diffuse into or out of the model region at a rate depending on the concentration difference.

\section{Parameter Values} \label{sec:parameterTables}

In \tabsref{tab:scotian_npzParameters}, \ref{tab:scotian_spratGlobalParameters}, and \ref{tab:scotian_spratSpeciesParameters}, we list the parameter values for our \ac{NPZ} model and the SPRAT model.

\begin{paramTable}
	\begin{tabular}{lll}
		\toprule
		Symbol          & Description                          & Value                                               \\ 
		\midrule
		$\gamma$        & Zooplankton assimilation efficiency  & $0.7$                                               \\
		$\iota_s$       & Half saturation irradiance           & $5$                  MJ m$^{-2}$                    \\
		$V_m$           & Uptake rate of nutrients             & $0.15$               d$^{-1}$                       \\
		$k_s$           & Half saturation nutrients            & $30$                 mmol N m$^{-2}$                \\
		$R_m$           & Uptake rate of phytoplankton         & $0.5$                d$^{-1}$                       \\
		$\lambda_s$     & Ivlev constant for grazing           & $\nicefrac{1}{600}$  (mmol N)$^{-1}$ m$^2$          \\
		$\varepsilon_P$ & Mortality rate phytoplankton         & $3.75 \cdot 10^{-4}$ (mmol N)$^{-1}$ m$^2$ d$^{-1}$ \\
		$\varepsilon_Z$ & Mortality rate zooplankton           & $5 \cdot 10^{-4}$    (mmol N)$^{-1}$ m$^2$ d$^{-1}$ \\
		$\varepsilon_f$ & Zooplankton mortality due to grazing & $0.075$              d$^{-1}$                       \\
		$\eta_r$        & Remineralization rate                & $24$                 y$^{-1}$                       \\
		$N_m$           & Maximum nutrient concentration       & $170$                mmol N m$^{-2}$                \\
		$C_c$           & Cloud coverage                       & $0.7$                                               \\ 
		\bottomrule
	\end{tabular}
	\caption{Parametrization of our \acs{NPZ} model for the eastern Scotian Shelf.}
	\label{tab:scotian_npzParameters}
\end{paramTable}

\begin{paramTable}
	\begin{tabular}{lll}
		\toprule
		Symbol          & Description                                                                 & Value         \\ 
		\midrule
		$\tau$          & Minimal predator\hyp{}prey mass ratio                                       & $8$           \\
		$r_{\tn{view}}$ & Perfect information radius (\autoref{eq:spratModel_informationSet})         & $10^5$ m      \\
		$\zeta_0$       & Scaling for feeding migration threshold  (\autoref{eq:spratModel_Vfeeding}) & $1$ kg$^{-1}$ \\ 
		\bottomrule
	\end{tabular}
	\caption{Parametrization of SPRAT for the eastern Scotian Shelf (global parameters).}
	\label{tab:scotian_spratGlobalParameters}
\end{paramTable}

\begin{paramTable} \hspace*{-6mm}
	\begin{tabular}{lllll>{\RaggedRight}p{4.1cm}}
		\toprule
		Symbol               & Description                      & Unit                        & Pred.                          & Forage                           & Source                                              \\ \midrule
		$C_{\tn{C/d}}$      & Carbon to dry mass ratio         & \du                         & $0.5$                          & $0.5$                            & \noparencite{watson2014}                            \\
		$C_{\tn{d/w}}$      & Dry to wet mass ratio            & \du                         & $0.25$                         & $0.25$                           & \ibid                                               \\
		$w_{\tn{egg}}$       & Egg dry mass                     & kg                          & $7 \cdot 10^{-8}$              & $1.4 \cdot 10^{-7}$              & \noparencite{ouellet2001}, \noparencite{hempel1967} \\
		$w_{\tn{forage}}$    & Min.~forage dry mass             & kg                          & $1.2 \cdot 10^{-7}$            & $2.43 \cdot 10^{-7}$             & \ibid                                               \\
		$M_{\tn{plankton}}$  & Max.~Z consumption wet mass      & kg                          & $0.05$                         & $r_{\max}$                       & \noparencite{jobling1981b}                          \\
		$M_{\tn{mature}}$    & Wet mass at maturity             & kg                          & $1.2$                          & $0.173$                          & \noparencite{froese2016}                            \\
		$M_{\tn{max}}$       & Maximum wet mass                 & kg                          & $14.5$                         & $0.732$                          & \ibid                                               \\
		$a$                  & $a$ for L-W relationship         & \du                         & $7.9 \cdot 10^{-3}$            & $6.9 \cdot 10^{-3}$              & \ibid                                               \\
		$b$                  & $b$ for L-W relationship         & \du                         & $3.05$                         & $3.04$                           & \ibid                                               \\
		$\varsigma$          & Cruise speed                     & BL s$^{-1}$                 & $0.6$                          & $0.8$                            & \noparencite{videler1991}                           \\
		$S$                  & Mating season                    & \na                         & $[\frac{2}{12}, \frac{5}{12}]$ & $[\frac{7}{12}, \frac{8.3}{12}]$ & \noparencite{froese2016}                            \\
		$\theta_{\tn{yolk}}$ & Duration till $w_{\tn{forage}}$  & s                           & $2.85 \cdot 10^{6}$            & $2.85 \cdot 10^{6}$              & \ibid                                               \\
		$E$                  & Assimilation efficiency          & \du                         & $0.8$                          & $0.8$                            & \noparencite{simenstad1986}                         \\
		$\varepsilon_B$      & Background mortality rate        & m$^2$ s$^{-1}$              & $1.47 \cdot 10^{-4}$           & $7.35 \cdot 10^{-5}$             & * \\
		$\zeta_B$            & Scaling of temperature mortality & s$^{-1}$ $^{\circ}$C$^{-1}$ & $5$                            & $0$                              & * \\
		$\Phi$               & Net wet mass fecundity           & kg$^{-1}$                   & $1.3 \cdot 10^{6}$             & $4 \cdot 10^{5}$                 & \noparencite{froese2016}                            \\
		$\eta$               & Predation rate                   & s$^{-1}$                    & $10^{-5}$                      & $10^{-5}$                        & * \\
		$F_0$                & Predation half sat.              & kg m$^{-2}$                 & $0.05$                         & $0.01$                           & * \\
		$\mu$                & Zooplankton grazing rate         & s$^{-1}$                    & $5.1 \cdot 10^{-8}$            & $5.5 \cdot 10^{-8}$              & * \\
		$Z_0$                & Zooplankton grazing half sat.    & kg m$^{-2}$                 & $5 \cdot 10^{-5}$              & $5 \cdot 10^{-5}$                & * \\ \bottomrule
	\end{tabular}
	\caption{Parametrization of SPRAT for the eastern Scotian Shelf (species parameters).}
	\label{tab:scotian_spratSpeciesParameters}
\end{paramTable}

\section*{References}

\bibliographystyle{plainnat}
\bibliography{main}

\end{document}